\documentclass{article}
\usepackage{graphicx}
\usepackage{amsmath,amssymb}
\usepackage{graphics,color,array,calc,rotating,epsfig,psfrag}
\usepackage{hyperref}
\numberwithin{equation}{section}
\usepackage{cite}
\usepackage{bm}
\usepackage{dcolumn}
\usepackage{float}
\usepackage{subfigure}
\usepackage{tikz-cd}
\usepackage{caption}
\usepackage[utf8]{inputenc}
\usepackage{tikz}
\usepackage{amsfonts}
\usepackage[mathscr]{eucal}
\usepackage{helvet}
\def\be{\begin{equation}} \def\ee{\end{equation}}
\def\bea{\begin{eqnarray}} \def\eea{\end{eqnarray}}

\newcommand{\nn}{\nonumber}
\newcommand{\sech}{\mathrm{sech}}
\newcommand{\RN}[1]{%
  \textup{\uppercase\expandafter{\romannumeral#1}}%
}
\begin{document}
	\baselineskip 18pt%
	\begin{titlepage}
		\vspace*{1mm}%
		\hfill%
		\vspace*{15mm}%
	
		\begin{center}
			{\Large {\bf \boldmath Generalized $T\overline{T}$-like Deformations in Duality-Invariant Nonlinear Electrodynamic Theories}}
		\end{center}
		\vspace*{5mm}
		\begin{center}
			{H. Babaei-Aghbolagh$^{1}$, Song He$^{2,3,4}$,
			 and Hao Ouyang$^{2}$
			}\\
			\vspace*{0.2cm}
			{\it
				$^{1}$Department of Physics, University of Mohaghegh Ardabili,
				P.O. Box 179, Ardabil, Iran\\
				$^{2}$Center for Theoretical Physics and College of Physics, Jilin University, 
				Changchun 130012, China\\
                $^{3}$School of Physical Science and Technology, Ningbo University, Ningbo, 315211, China\\
				$^{4}$Max Planck Institute for Gravitational Physics (Albert Einstein Institute),\\
				Am M\"uhlenberg 1, 14476 Golm, Germany\\
			}
			
			\vspace*{0.5cm}
			{E-mails: {\tt h.babaei@uma.ac.ir,  hesong@jlu.edu.cn, haoouyang@jlu.edu.cn
			}}
			\vspace{1cm}
		\end{center}
		
		\begin{abstract}

%In this paper, we present a high-order perturbation approach to classify two types of general solutions for the duality-invariant nonlinear electrodynamic theories that satisfy the differential self-duality condition. {\color{red}  These solutions encompass the large class of duality-invariant electrodynamic theories. } One type corresponds to irrelevant stress tensor $T\bar{T}$-like flows, while the other type involves combinations of irrelevant $T\bar{T}$-like and marginal root-$T\bar{T}$-like deformations. Our framework enables us to explore various duality-invariant nonlinear electrodynamics theories and their associated irrelevant and marginal stress tensor flows. Additionally, we demonstrate the commutativity for the flows driven by the irrelevant and marginal operators.

This study introduces a high-order perturbation methodology to categorize two primary solution types within duality-invariant nonlinear electrodynamic theories, adhering to the differential self-duality criterion. The first solution type aligns with irrelevant stress tensor flows, resembling $T\bar{T}$ dynamics, and the second involves a blend of irrelevant $T\bar{T}$-like and marginal root-$T\bar{T}$-like deformations. Our approach facilitates the investigation of diverse duality-invariant nonlinear electrodynamics theories and their stress tensor flows and confirms the commutativity of flows initiated by irrelevant and marginal operators.

		\end{abstract}	
		
	\end{titlepage}
	
%%%%%%%%%%%%%%%%%%%%%%%%%%%%%%%%%%%%%%%%%%%%%%%%%%%
\section{Introduction}\label{0}
%%%%%%%%%%%%%%%%%%%%%%%%%%%%%%%%%%%%%%%%%%%%%%%%%%%%

A fundamental concept in nonlinear electrodynamics (NED) is adding interaction terms to Maxwell's theory to preserve $SO(2)$ duality symmetry, enhancing our understanding of NEDs. There are three principal methods to construct self-dual theories of Maxwell's fields with $SO(2)$ duality symmetry: the Gaillard-Zumino approach \cite{Gaillard:1981rj, Gaillard:1997rt}, improved by Gibbons and Rasheed \cite{Gibbons:1995ap}; the non-covariant Hamiltonian approach by Henneaux and Teitelboim \cite{HT}, followed by Deser et al. \cite{Deser:1997mz, Deser:1997se}; and the PST method by Pasti, Sorokin, and Tonin \cite{Pasti:1995tn, Pasti:1995us, Pasti:1996vs, Pasti:1997gx}. These methods offer different frameworks for incorporating interaction terms while preserving $SO(2)$ duality symmetry. This paper presents a method for constructing effective Lagrangians for nonlinear theories that respect duality symmetries, building on irrelevant $ T\bar{T} $-like and marginal root-$ T\bar{T} $-like deformations in NED theories \cite{Conti:2018jho, Babaei-Aghbolagh:2022MoxMax}.

%It aims to enhance our understanding of counterterms and the $ E_{7(7)} $ electromagnetic duality in $ {\cal N}=8 $ supergravity \cite{Kallosh:2011dp, Bossard:2011ij, Carrasco:2011jv}. Focusing on a model with a single vector field $ F_{\mu \nu} $, it revisits the unique features of $ E_{7(7)} $ symmetry, first identified by Gaillard and Zumino, and the nuances of duality in the vector sector \cite{Gaillard:1981rj, Gaillard:1997rt}. In Maxwell's theories, nonlinear deformations are optional due to their inherent lack of interaction. Conversely, supergravity theories often require modifications guided by counterterms from precise calculations. These adjustments, though not the only factors for duality alignment, are crucial for developing an action that respects duality symmetry \cite{Carrasco:2011jv, Chemissany:2011yv}. The discussion spans U(1) duality theories from Maxwell's electromagnetism to nonlinear Born-Infeld theory, emphasizing the latter's duality and significance in string theory.

Using the Gaillard-Zumino approach, we obtain the general form of electrodynamic theories with duality invariance.
The Lagrangian of  duality invariant NED theories satisfy the self-duality condition $G\tilde{G}+F\tilde{F}=0$, \cite{Bialynicki-Birula:1981, Bialynicki-Birula:1992rcm}(see also \cite{Gibbons:1995ap, Gaillard:1997rt, Kuzenko:2000uh, Aschieri:2008ns,Kallosh:2011dp,Bossard:2011ij,Carrasco:2011jv,Chemissany:2011yv,Aschieri:2013nda}) where $\tilde{G}$ is an antisymmetric tensor field given by
$\tilde{G}^{\mu \nu} = 2 \frac{ \partial {\cal L}(F)}{\partial F_{\mu \nu}}\,$
and the factor 2 accounts for the antisymmetry of ${F}_{\mu\nu}$.
It can be easily shown that
$
\tilde G_{\mu \nu} =  \mathcal{L}_t F_{\mu \nu} + \mathcal{L}_z \tilde F_{\mu \nu}$ and $ G_{\mu \nu} = - \mathcal{L}_t\tilde F_{\mu \nu}+ \mathcal{L}_z F_{\mu \nu},
$
  where we have used the standard notation $ \mathcal{L}_t=\frac{ \partial {\cal L}}{\partial t}$ and $\mathcal{L}_z=\frac{ \partial {\cal L}}{\partial z}$, which  $ t=\frac{1}{4}F_{\mu\nu}F^{\mu\nu}$ and $\,\,z=\frac{1}{4}F_{\mu\nu}\tilde F^{\mu\nu}$ are two Lorentz invariant variables.
The differential form of the self-duality condition is:
\begin{eqnarray}
\label{lagrangeNGZ}
\big( (\partial_t \mathcal{L})^2-(\partial_z \mathcal{L})^2-1\big)\,z -\big(2 \, (\partial_z \mathcal{L}) (\partial_t \mathcal{L})\big)\, t=0 \,.
\end{eqnarray}
The self-dual condition in Eq. \ref{lagrangeNGZ} has multiple solutions for the NED Lagrangians. Ref. \cite{cour} describes the general methods to solve the differential Eq. \ref{lagrangeNGZ}, and Refs. \cite{Kuzenko:2000uh, Carrasco:2013qia} explain how to solve its perturbation. Additionally, innovative approaches can be explored for resolving the self-duality differential equation in Refs. \cite{Mkrtchyan:2205uvc, russo2024causal}.

In the Gaillard-Zumino approach, Maxwell's field equations have an electric-magnetic $SO(2)$ duality symmetry, but extending this symmetry to the action level is challenging. This becomes more difficult for theories with nonlinear interactions of Maxwell's fields. The Lagrangians are not $SO(2)$ duality-symmetric, but their derivatives with respect to the invariant parameters are. The invariant parameter can be a coupling constant or a background field, like the gravitational field, that is duality-rotation invariant\cite{Gaillard:1981rj}.
In this approach, the self-dual condition implies that the physical objects of the theory, such as equations of motion\cite{Green:1996qg}, energy-momentum tensor\cite{Gibbons:1995ap,BabaeiVelni:2016qea} and scattering amplitudes\cite{Babaei-Aghbolagh:2013hia,Garousi:2017fbe,Pavao2210be,BabaeiVelni:2019ptj}, are S-dual invariant. In the PST approach, incorporating auxiliary fields, various formulations of Maxwell's theory action can be 
constructed to exhibit duality symmetry \cite{Mkrtchyan:2019opf}.

A recent development in 2D field theory is the addition of an irrelevant operator $O_\lambda$ to the free Lagrangian, which results in a deformed Lagrangian of the form:
\begin{equation}
\label{L1}
{\cal L}_{\lambda}={\cal L}_{free} +\int O_\lambda d\lambda ,
\end{equation}
where $\lambda$ is a dimensionful coupling parameter, the deformation operator $O_{\lambda}$ is a special function of the energy-momentum tensor of the free Lagrangian, which can be written as:
$
O_{\lambda}=\frac{1}{2} \Big(  {T_{\mu\nu}T^{\mu\nu}-T_{\mu}}^{\mu} {T_{\nu}}^{\nu}\Big)\,.
$  
The operator $O_{\lambda}$ is known as the $T \bar{T}$ deformation operator\cite{Smirnov:2016lqw}\cite{Cavaglia:2016oda}, and it involves the energy-momentum tensor, which is a symmetric tensor of the standard form:
$
\label{eq:Hstressen}
T_{\mu\nu}=\dfrac{-2}{\sqrt{g}}\dfrac{\delta{\cal L}}{\delta g^{\mu\nu}}
$.
A similar approach can be applied in four dimensions. Maxwell's theory is treated as a free theory, and additional interaction terms are introduced to Maxwell's free Lagrangian using an operator similar to two dimensions with a coupling constant $\lambda$. By choosing the deformation operator to be \begin{equation}O_{\lambda}=\frac{1}{8} \Big(T_{\mu\nu}T^{\mu\nu}  -\frac{1}{2} {T_{\mu}}^{\mu} {T_{\nu}}^{\nu}\Big),
\end{equation}
the extra terms in Maxwell's theory correspond precisely to the expansion of the Born-Infeld Lagrangian:
\begin{equation}
\label{LEBI}
{\cal L}_{BI}={\cal L}_{Max} +\int O_\lambda d\lambda =- t + \tfrac{1}{2} \lambda (t^2 + z^2) -  \tfrac{1}{2} t \lambda^2 (t^2 + z^2)+\dots\,\,.
\end{equation}
The Born-Infeld Lagrangian theory satisfies the differential self-duality condition in Eq. \ref{lagrangeNGZ}. It has the following closed-form expression:
\begin{equation}
{\cal L}_{BI }  =\frac{1}{\lambda} \bigg[ 1 -  \sqrt{1+ 2 \lambda  t-\lambda^2 z^2 } \bigg].
\end{equation}
 One of our interests is to find the general form of $T\bar{T}$-like deformations in NED theories. In Refs. \cite{Conti:2018jho, Babaei-Aghbolagh:2020kjg,Aghbolagh2210,Ferko2023}, proposed some methods to study the relation between duality invariant NED theories and the $T\bar{T}$-like deformations.
 As demonstrated in Ref. \cite{Conti:2018jho}, the Born-Infeld theory prescribes a flow equation given by: $\frac{\partial \mathcal{L}{BI}}{\partial \lambda} = \frac{1}{8} \left( T_{\mu \nu} T^{\mu \nu} - \frac{1}{2} {T_{\mu}}^{\mu} {T_{\nu}}^{\nu} \right),$ notably, this is not the sole example of a duality invariant NED theory.
 There are many other duality invariant theories. The Bossard-Nicolai theory stands as another renowned example within the realm of duality-invariant NEDs~\cite{Bossard:2011ij, Carrasco:2011jv}. In the framework of our general auxiliary-field approach to the 
duality-invariant models, it was called there ``{\it the simplest interaction model}''~\cite{Ivanov:2004jv,Ivanov:2013jv}.This paper found a non-trivial flow equation for the Bossard-Nicolai theory. By generalizing this approach, we obtained the general form of the flow equation in duality invariant NED theories. From this general flow equation, we can derive more non-trivial flow equations.

One of the motivations of the present work comes from the advancements in non-linear electrodynamics, particularly the ModMax electrodynamics\cite{Bandos:2020jsw}. This theory extends Maxwell's electrodynamics into the non-linear domain while retaining SO(2) and conformal invariance. The ModMax Lagrangian is expressed as:
\begin{equation}
\label{LMM}
{\cal L}_{MM}=-\cosh\gamma \,t+\sinh \gamma\,\sqrt{t^2+z^2}.
\end{equation}
The unitarity condition, $\gamma \geq 0$, allows for exact solutions resembling Maxwell's plane waves with arbitrary polarization. The ModMax theory emerges from the dimensional reduction of chiral 2-form electrodynamics in six dimensions, adhering to the self-duality condition \cite{Bandos:2020hgy, Deger2024cw}. Unlike Maxwell's electrodynamics, which is the weak-field limit, ModMax maintains conformal invariance and reverts to Maxwell's theory as the parameter $\gamma$ approaches zero. Further research on this topic is detailed in  Refs.~\cite{Kosyakov:2020wxv,Bandos:2021rqy,Kuzenko:2021cvx,Avetisyan:2021heg,Bansal:2021bis,Kruglov:2021bhs,BallonBordo:2020jtw,Bakhtiarizadeh:2023mhk,Nastase:2021uvc,Kuzenko2024c,Christian2024cw}, providing a deeper exploration of the theory's nuances and implications.

The ModMax theory has a marginal deformation parameter that preserves the conformal symmetry and the gauge invariance of Maxwell's theory. In a recent work \cite{Babaei-Aghbolagh:2022MoxMax}, one of the authors and his collaborators introduced the ModMax theory as a root-type $ T \bar{T} $ deformation of Maxwell's theory.
The root-type $T \bar{T}$ operator in four dimensions is:
\begin{eqnarray}\label{Ogama}
{\cal R}_{\gamma}= \frac{1}{2}
\sqrt{T_{\mu\nu}T^{\mu\nu}- \frac{1}{4} {T_{\mu}}^{\mu} {T_{\nu}}^{\nu}}\,.
\end{eqnarray} 
Ref. \cite{Babaei-Aghbolagh:2022MoxMax} demonstrates that a root-type $ T \bar{T} $-like deformation facilitates the transformation of the BI theory into the Generalized Born-Infeld (GBI) theory (a BI-type deformation of ModMax). The GBI theory is a generalization of the BI theory that depends on two Lorentz scalars $t$ and $z$ in $D=4$, and its Lagrangian density is given by:
\begin{equation}\label{GBI1}
{\cal S}_{GBI}(\lambda ,\gamma) =\int \frac{1}{\lambda} \Bigg[ 1 -  \sqrt{1 -  \lambda \Bigl( 2
	\big( - \cosh(\gamma)t+\sinh(\gamma) \sqrt{ t^2+z^2} \big) +\lambda z^2 \Bigr)} \Bigg]d^dx .\nn
\end{equation}
The GBI theory is a solvable deformation of Maxwell's and BI theories that depends on two parameters $\gamma$ and $\lambda$. It obeys the flow equations $\partial {\cal L}_{GBI}/{\partial \gamma} = {\cal R}_{\gamma}$ and  $\partial {\cal L}_{GBI}/{\partial \lambda} = O _{\lambda}$, where ${\cal R}_{\gamma}$ and $O _{\lambda}$ are the root-type and irrelevant  $T \bar{T}$-like operators in four dimensions. We have demonstrated that these operators commute with each other, as shown in Fig. \ref{fig}.
\begin{center}
	\begin{tikzcd}
	{\cal S}_{Max} \arrow[r, blue, "O_{\lambda}" blue] \arrow[d,red,"{\cal R}_{\gamma}" red]
	&|[blue]| {\cal S}_{BI}(\lambda) \arrow[d,red, "{\cal R}_{\gamma}" red] \\
	|[red]|{\cal S}_{MM }(\gamma) \arrow[r, blue, "O_{\lambda}" blue]
	&|[red!50!blue]|  {\cal S}_{GBI}( \lambda, \gamma)
	\end{tikzcd}
	\captionof{figure}{A commutative diagram  between operators ${\cal R}_{\gamma}$ and $O_{\lambda}$ non-linear electrodynamic theories.}\label{fig}
\end{center}
 A two-dimensional version of the ModMax theory was investigated in the\cite{Conti:2022egv}, where the authors in the second version introduced the root-type $ T \bar{T} $ deformation associated with ModMax theory in two dimensions. In two dimensions, root-type $ T \bar{T} $ deformation was independently discovered in the \cite{Babaei-Aghbolagh:2022kjj, Ferko:2206jsw}. Root-type $ T \bar{T} $ deformation  has been developed for various theories in different dimensions \cite{Hou:2022csf,Borsato:2022tmu,Garcia:2022wad,Tempo:2022ndz, Aramini:2022wbn,Ferko:2023ruw,Ferko:2023sps,Ebert:2023tih,Rodriguez:2021tcz,Bagchi:2022tcz,Ferko:2023ozb,Ferko:2023iha,he2024irrelevan,Bagchi:2024cz,ferko2024interacting,Ebert:2024zwv,Tian2024c,Babaei2024c,Chen2024c,Ferko2024cw,Tsolakidis2024cw, Hadasz2024cw}.

We investigate the solutions of the differential Eq. \ref{lagrangeNGZ} with the self-dual condition and two types of perturbation solutions. The first type has a dimensionful $\lambda$ coupling, and we take it as a function of the integers power of $t$ and $z$ in the form of $\mathcal{L}(\lambda)=\mathcal{F}(t^N,z^M,\lambda^{N+M-1})$. The second type has two coupling parameters, one dimensionful $\lambda$ and one dimensionless $\gamma$. Besides the integers power of $t$ and $z$, this type also has the square root of these two Lorentz derivatives in the form of $\mathcal{L}(\lambda, \gamma)=\mathcal{G}(t^N,z^M,{(\sqrt{t^2+z^2})}^q,\lambda^{N+M+q-1},\gamma)$ where $q$ is equal to 1 or 0. Our goal is to classify self-dual NED theories. These theories have two kinds of coupling parameters: a dimensionless one ($\gamma$) and a dimensionful one ($\lambda$). We will demonstrate how to derive the expressions for $\gamma$ and $\lambda$ couplings from root-type and irrelevant  $ T \bar{T} $ deformations, respectively.

The organization of this paper is as follows: In section \ref{02}, we explore the elevated orders of $ \lambda $, incorporating integer powers of the variables $ t $ and $ z $ to ensure the Lagrangian aligns with the self-duality condition. Our findings indicate that at these higher orders of $ \lambda $, the Lagrangians exhibit a general irrelevant flow equation.
We particularly verify the nature of this irrelevant flow equation in the context of Born-Infeld and Bossard-Nicolai actions. In section \ref{Secdn3}, we examine the general Lagrangian introduced previously, now incorporating the term $ \sqrt{t^2 + z^2} $. This addition requires the Lagrangian's coefficients to be linked through a dimensionless coupling constant $ \gamma $. These coefficients are derived to comply with the self-duality condition and adhere to a general marginal flow equation. We demonstrate that, in a particular instance, this general marginal flow equation simplifies to the root marginal flow equation and is commuted with the irrelevant flow equation. Finally, in section \ref{04}, we provide a summary of our results and outlook.

%%%%%%%%%%%%%%%%%%%%%%%%%%%%%%%%%%%%%%%%%%%%%%%%%%%
%%%%%%%%%%%%%%%%%%%%%%%%%%%%%%%%%%%%%%%%%%%%%%%%%%%%%%%%%%%%%%%%%%%%%%%%%%%%%%%%%%%%%%%%%%%%%%%%%%%%%%%%%%%%%%%%%%%%%%%%%%%%%%%%%%%%%%%%%%%%%%%%%%%%%%%%%%%%%%%%%%%%%%%%%%%%%%%%%%%%%%%%%%%%%%%%%%%%%%%%%%%%%%%%%%%%%%%%%%%%%%%%%%%%%%%%%%%%%%%%%%%%%%%%%%%%%%%%%%%%%%%%%%%%%%%%%%%%%%%%%%%%%%%%%%%%%%%%%%%%
\section{Pertubative self-dual solutions with irrelevant deformations}\label{02}
%%%%%%%%%%%%%%%%%%%%%%%%%%%%%%%%%%%%%%%%%%%%%%%%%%%%%%%%%%%%%%%%%%%%%%%%%%%%%%%%%%%%%%%%%%%%%%%%%%%%%%%%%%%%%%%%%%%%%%%%%%%%%%%%%%%%%%%%%%%%%%%%%%%%%%%%%%%%%%%%%%%%%%%%%%%%%%%%%%%%%%%%%%%%%%%%%%%%%%%%%%%%%%%%%%%%%%%%%%%%%%%%%
The self-duality condition in Eq.~\ref{lagrangeNGZ} is necessary for the Lagrangian of a nonlinear electromagnetic theory that preserves duality symmetry. In this section, we employ a perturbative method to address the self-duality condition, utilizing a series expansion in integer powers of the Lorentz variables $ t $ and $ z $ with a dimensionful coupling parameter $\lambda$. We consider a general Lagrangian expressed as a series encompassing all orders of $\lambda$. For each power of $\lambda$, we explore every conceivable combination of integer powers of $ z $ and $ t $. This general Lagrangian is given by:
\begin{eqnarray}
\label{lagrangeGeneral}
\mathcal{L}(\lambda^n)&=&a_1 t + \lambda (a_2 t^2 +a_3 t z + a_4 z^2 )+ \lambda^2 (a_5 t^3 +a_6 t^2 z + a_7 t z^2 +a_8 z^3)\nonumber\\
&&+\lambda^3 (a_9 t^4 + a_{10} t^3 z + a_{11} t^2 z^2 + a_{12} t z^3 + a_{13 } z^4) \nonumber\\
&& +\lambda^4 (a_{14} t^5 + a_{15} t^4 z + a_{16} t^3 z^2 + a_{17} t^2 z^3 + a_{18 } t z^4 + a_{19} z^5)\nonumber\\
&& +\lambda^5 (a_{20} t^6 + a_{21} t^5 z + a_{22} t^4 z^2 + a_{23} t^3 z^3 + a_{24} t^2 z^4 + a_{25} t z^5 + a_{26} z^6)+\dots\nonumber\\
&&+ \lambda^n (a_{m-n} t^{n+1}+a_{m-n+1} t^{n} z+\dots.+ a_{m-1} t^2 z^{n-1} +a_{m} t z^n +a_{m+1} z^{n+1} )+...
\end{eqnarray}
where  $m=\sum_{i=1}^{n} (i+2)$ for $n \geq 2$ and  $n=N+M-1$. The Lagrangian detailed in Eq. \ref{lagrangeGeneral} encapsulates the general structure for all irrelevantly deformed Lagrangians characterized by integer powers of $t$ and $ z$. A systematic analysis of Eq. \ref{lagrangeNGZ}, applied iteratively to Lagrangian \ref{lagrangeGeneral}, reveals that the $a_m$ coefficients must conform to specific constraints. These constraints are essential for the differential equation's solution. Therefore, these coefficients must adhere to the stipulated conditions up to an order of $ \lambda^7$:
\begin{eqnarray}
\label{an}
&&\Big(a_{1 }\to -1\Big)\,\,\,\,\,;\,\,\,\,\,\, \Big(a_{3 }\to 0\,\,,\,\,\, a_{4} \to a_{2}\Big) \,\,\,\,\,;\,\,\,\,\,\, \Big(a_{5} \to -2 a_{2}^2\,\,,\,\,\, a_{6} \to -2 a_{2}^2\,\,,\,\,\,
a_{7}=a_{8} \to 0\Big)\,\,\,\,\,;\,\,\,\,\,\,\nonumber\\
&& \Big(a_{10 }= a_{12 }\to 0\,\,,\,\,\, 
a_{13 }\to 1/2 (a_{11} - 4 a_{2}^3)\,\,,\,\,\, 
a_{9} \to 1/2 (a_{11} + 4 a_{2}^3)\Big) \,\,\,\,\,;\,\,\,\,\,\,\nonumber\\
&& \Big(a_{14} \to -2 (2 a_{11} a_{2 }- 5 a_{2}^4)\,\,,\,\,\, 
a_{15}=a_{17}=a_{19} \to 0\,\,,\,\,\, a_{16} \to -4 (2 a_{11} a_{2} - 7 a_{2}^4)\,\,,\,\,\,
a_{18} \to -2 (2 a_{11} a_{2 }- 9 a_{2}^4)\Big) \,\,\,\,\,;\,\,\,\,\,\,\nonumber\\
&& \Big(a_{20} \to 
1/3 (20 a_{11} a_{2}^2 - 64 a_{2}^5 + a_{22})\,\,,\,\,\, a_{21} = a_{23}= a_{25} \to 0\,\,,\,\,\, \nonumber\\
&&a_{24} \to -20 a_{11} a_{2}^2 + 80 a_{2}^5 + a_{22}\,\,,\,\,\, 
a_{26} \to 1/3 (-40 a_{11} a_{2}^2 + 176 a_{2}^5 + a_{22})\Big) \,\,\,\,\,;\,\,\,\,\,\,\nonumber\\
&& \Big(a_{27} \to 
1/16 ( 208 a_{11} - 32 a_{11}^2 - 32 a_{22}-101 )\,\,,\,\,\, a_{28}=a_{30} \to 0\,\,,\,\,\, 
a_{29} \to 1/16 ( 784 a_{11} - 96 a_{11}^2 - 96 a_{22}-387 )\,\,,\nonumber\\
&&  a_{31} \to 1/16 ( 944 a_{11} - 96 a_{11}^2 - 96 a_{22}-479 )\,\,,\,\,\, a_{32}=a_{34} \to 0\,\,,\,\,\, 
a_{33} \to 1/16 ( 368 a_{11} - 32 a_{11}^2 - 32 a_{22}-193 )\Big) \,\,;\,\,\,\,\,\,\nonumber\\
&& \Big(a_{35} \to 1/4 (25 - 56 a_{11} + 14 a_{11}^2 + 7 a_{22} + a_{37})\,\,,\,\,\, 
a_{39} \to 1/4 ( 406 a_{11} - 84 a_{11}^2 - 42 a_{22} + 6 a_{37}-189 )\,\,,\nonumber\\
&&  
a_{36}= a_{38}= a_{40}= a_{42} \to 0\,\,,\nonumber\\
&& 
a_{41} \to -70 + 147 a_{11} - 28 a_{11}^2 - 14 a_{22} + a_{37}\,\,,\,\,\, 
a_{43} \to 1/4 (-116 + 238 a_{11} - 42 a_{11}^2 - 21 a_{22} + a_{37})\Big).
\end{eqnarray}
Using the conditions \ref{an} on the coefficients of $\mathcal{L}(\lambda^n)$, we get the following Lagrangian that is duality invariant up to order $\lambda^7$:
%\begin{eqnarray}
% \label{Ltz1}
% {\cal L}(\lambda^n)&=& - t + \tfrac{1}{2} \lambda (t^2 + z^2) -  \tfrac{1}{2} t \lambda^2 (t^2 + z^2) + \tfrac{1}{4} \lambda^3 (t^2 + z^2) \bigl((1 + 2 a_{11}) t^2 + ( 2 a_{11}-1) z^2\bigr)\nonumber\\
% && -  \tfrac{1}{8} t \lambda^4 (t^2 + z^2) \bigl(( 16 a_{11}-5) t^2 + (16 a_{11}-9) z^2\bigr)\nonumber\\
% && + \lambda^5 \bigl(\tfrac{1}{3} ( 5 a_{11} + a_{22}-2) t^6 + a_{22} t^4 z^2 + (\tfrac{5}{2} - 5 a_{11} + a_{22}) t^2 z^4 + \tfrac{1}{6} (11 - 20 a_{11} + 2 a_{22}) z^6\bigr)\nonumber\\
% && -  \tfrac{1}{16} t \lambda^6 (t^2 + z^2) \Bigl(\bigl(101 + 16 a_{11} ( 2 a_{11}-13 ) + 32 a_{22}\bigr) t^4 + 2 \bigl(143 + 32 ( a_{11}-9 ) a_{11} + 32 a_{22}\bigr) t^2 z^2 \nonumber\\
% &&+ \bigl(193 + 16 a_{11} ( 2 a_{11}-23) + 32 a_{22}\bigr) z^4\Bigr)\nonumber\\
% && + \tfrac{1}{4} \lambda^7 \biggl(\bigl(25 + 14 ( a_{11}-4) a_{11} + 7 a_{22} + a_{37}\bigr) t^8 + 4 a_{37} t^6 z^2 \nonumber\\
% &&+ \Bigl(-7 \bigl(27 + 2 a_{11} (-29 + 6 a_{11}) + 6 a_{22}\bigr) + 6 a_{37}\Bigr) t^4 z^4\\
% && + 4 \bigl(7 (21 - 4 a_{11}) a_{11} - 14 (5 + a_{22}) + a_{37}\bigr) t^2 z^6 + \bigl( 14 (17 - 3 a_{11}) a_{11} - 21 a_{22} + a_{37}-116 \bigr) z^8\biggr)\,,\nonumber
% \end{eqnarray}
\begin{eqnarray}
\label{Ltz}
{\cal L}(\lambda^n)&=&- t + \tfrac{1}{2} \lambda (t^2 + z^2) -  \tfrac{1}{2} t \lambda^2 (t^2 + z^2)+ \lambda^3 \bigl(\tfrac{1}{2} a_{11} (t^2 + z^2)^2 + \tfrac{1}{4} (t^4 -  z^4)\bigr)\nonumber\\
&& + \lambda^4 \bigl(-2 a_{11} t (t^2 + z^2)^2 + \tfrac{1}{8} t (t^2 + z^2) (5 t^2 + 9 z^2)\bigr) \nonumber\\
&& + \lambda^5 \bigl(\tfrac{1}{3} a_{22} (t^2 + z^2)^3 + \tfrac{5}{3} a_{11} (t^6 - 3 t^2 z^4 - 2 z^6) + \tfrac{1}{6} (-4 t^6 + 15 t^2 z^4 + 11 z^6)\bigr)\nonumber\\
&& + \lambda^6 \bigl(-2 a_{11}^2 t (t^2 + z^2)^3 - 2 a_{22} t (t^2 + z^2)^3 + a_{11} t (t^2 + z^2)^2 (13 t^2 + 23 z^2)\nonumber\\
&& -  \tfrac{1}{16} t (t^2 + z^2) (101 t^4 + 286 t^2 z^2 + 193 z^4)\bigr)\nonumber\\
&&+ \lambda^7 \bigl(\tfrac{25}{4} t^8 -  \tfrac{189}{4} t^4 z^4 - 70 t^2 z^6 - 29 z^8 + \tfrac{7}{2} a_{11}^2 (t^2 - 3 z^2) (t^2 + z^2)^3\nonumber\\
&& + \tfrac{7}{4} a_{22} (t^2 - 3 z^2) (t^2 + z^2)^3 + \tfrac{1}{4} a_{37} (t^2 + z^2)^4 -  \tfrac{7}{2} a_{11} (t^2 + z^2)^2 (4 t^4 - 8 t^2 z^2 - 17 z^4)\bigr)+\dots
\end{eqnarray}
Assuming $a_2=1/2$, we have three unknown coefficients $( a_{11},a_{22},a_{37})$ up to order $\lambda^7$. For each order $\lambda^{2 n}$, we have $n$ unknown coefficients that can vary for different self-duality invariant theories. The appendix shows the duality invariant Lagrangian up to order $\lambda^{12}$ in Eq. \ref{Ap1}.
%%%%%%%%%%%%%%%%%%%%%%%%%%%%%%%%%%%%%%%%%%%%%%%%%%%%%%%%%%%%%%%%%%%%%%%%%%%%%%%%%%%%%%%%%%%%%%%%%%%%%%%%%%%%%5
\subsection{A non-trivial $T\bar{T}$-like deformation for Bossard-Nicolai Theory}
%%%%%%%%%%%%%%%%%%%%%%%%%%%%%%%%%%%%%%%%%%%%%%%%%%%%%%%%%%%%%%%%%%%%%%%%%%%%%%%%%%%%%%%
We consider the Lagrangian in Eq.~\ref{Ltz}, which satisfies both the duality condition and a flow equation involving the energy-momentum tensor. A notable example of a duality-invariant Lagrangian is the Born-Infeld Lagrangian. We can derive it from the general Lagrangian in Eq.~\ref{Ltz} by setting the coefficients $a_{11} = \frac{3}{4}$, $a_{22} = \frac{35}{16}$, and $a_{37} = \frac{231}{32}$ up to order $\lambda^7$.
%\begin{eqnarray}
%\label{BILag}
% {\cal L}_{\rm BI}= \frac{1}{\lambda} (1 - \sqrt{1 + 2 \lambda t - \lambda^2 z^2\,}) \,,
% \end{eqnarray}
The $T \bar{T}$-like flow equation is valid for the Born-Infeld theory at all orders of $\lambda$:
\begin{equation}\label{lM}
\frac{\partial {\cal L}_{BI}}{\partial \lambda}=\frac18 \left( T_{\mu\nu}T^{\mu\nu}-\frac{1}{2}  {T_{\mu}}^{\mu} {T_{\nu}}^{\nu}\right) \,.
\end{equation}
Besides the Born-Infeld theory, there exist other non-linear electrodynamic theories that preserve duality symmetry. One such solution of the differential equation \ref{lagrangeNGZ} is the Bossard-Nicolai theory in Refs.\cite{Bossard:2011ij, Carrasco:2011jv}, which follows from the general Lagrangian \ref{Ltz} by choosing $( a_{11}=1, a_{22}=\frac{177}{32}, a_{37}=\frac{1243}{32})$. The Lagrangian density of this theory depends on two Lorentz scalars $t$ and $z$ as follows
\begin{eqnarray}\label{BN}
\mathcal{L}_{BN}&=&- t + \tfrac{1}{2} \lambda (t^2 + z^2) -  \tfrac{1}{2} t \lambda^2 (t^2 + z^2) + \tfrac{1}{4} \lambda^3 (t^2 + z^2) (3 t^2 + z^2) -  \tfrac{1}{8} t \lambda^4 (t^2 + z^2) (11 t^2 + 7 z^2)\nonumber\\
&& + \tfrac{1}{32} \lambda^5 (t^2 + z^2) (91 t^4 + 86 t^2 z^2 + 11 z^4) -  \tfrac{1}{8} t \lambda^6 (t^2 + z^2) (51 t^4 + 64 t^2 z^2 + 17 z^4)\nonumber\\
&& + \tfrac{1}{64} \lambda^7 (t^2 + z^2) (969 t^6 + 1517 t^4 z^2 + 623 t^2 z^4 + 43 z^6)+{\cal O} ( \lambda^8).
\end{eqnarray}
We ask whether the Bossard-Nicolai theory shares the similar $T \bar{T}$-like flow equation as the Born-Infeld theory. Alternatively, what is the analog of the $T \bar{T}$-like flow equation \ref{lM} for the Bossard-Nicolai theory? To answer this question, we compute the energy-momentum tensor as $ T_{\mu\nu}=g_{\mu\nu} {\cal L}_{BN}-F_{\mu}{}^\rho \tilde{G}_{\nu \rho}$. We also look for a function of the energy-momentum tensor of the form $f(T_{\mu\nu},\lambda)$ that satisfies the flow equation $\frac{\partial {\cal L}_{BN}}{\partial \lambda}=f(T_{\mu\nu},\lambda)$.
Since the Bossard-Nicolai and Born-Infeld Lagrangians agree up to $\lambda^2$, the first term in the expansion of $f(T_{\mu\nu},\lambda)$ is $f (T_{\mu\nu},\lambda)=\tfrac{1}{8} T_{\mu\nu}T^{\mu\nu}+\dots$. The energy-momentum tensor of the Bossard-Nicolai Lagrangian is given by
% \begin{eqnarray}\label{GBN2}
% \tilde{G}_{\mu\nu}&=&-F_{\mu\nu } + \lambda ( t F_{\mu\nu } +  z \tilde{F}_{\mu\nu }) - \lambda^2 \bigl(\tfrac{1}{2} F_{\mu\nu } (3 t^2 + z^2) + t z \tilde{F}_{\mu\nu }\bigr)+ \lambda^3 \bigl( F_{\mu\nu } (3 t^3 + 2 t z^2) +  z (2 t^2 + z^2) \tilde{F}_{\mu\nu }\bigr)\nonumber\\
% && - \tfrac{1}{8} \lambda^4 \bigl(F_{\mu\nu } (55 t^4 + 54 t^2 z^2 + 7 z^4) + 4 t z (9 t^2 + 7 z^2) \tilde{F}_{\mu\nu }\bigr)\nonumber\\
% && +\tfrac{1}{16} \lambda^5 \bigl( F_{\mu\nu } (273 t^5 + 354 t^3 z^2 + 97 t z^4) +  z (177 t^4 + 194 t^2 z^2 + 33 z^4) \tilde{F}_{\mu\nu }\bigr) \nonumber\\
% &&- \tfrac{1}{8} \lambda^6 \bigl(F_{\mu\nu } (357 t^6 + 575 t^4 z^2 + 243 t^2 z^4 + 17 z^6) + 2 t z (115 t^4 + 162 t^2 z^2 + 51 z^4) \tilde{F}_{\mu\nu }\bigr)\\
% && +\tfrac{1}{16} \lambda^7 \bigl( F_{\mu\nu } (1938 t^7 + 3729 t^5 z^2 + 2140 t^3 z^4 + 333 t z^6) +  z (1243 t^6 + 2140 t^4 z^2 + 999 t^2 z^4 + 86 z^6) \tilde{F}_{\mu\nu }\bigr),\nonumber
% \end{eqnarray}
\begin{eqnarray}\label{TBN}
T_{\mu\nu}&=&T^{Max}_{\mu \nu } + \lambda \bigl(- t T^{Max}_{\mu \nu } -  \tfrac{1}{2} (t^2 + z^2) \mathit{g}_{\mu \nu }\bigr) + \lambda^2 \bigl(\tfrac{1}{2} T^{Max}_{\mu \nu } (3 t^2 + z^2) + t (t^2 + z^2) \mathit{g}_{\mu \nu }\bigr)  \nonumber\\
&&+ \lambda^3 \bigl(- T^{Max}_{\mu \nu } (3 t^3 + 2 t z^2) -  \tfrac{3}{4} (t^2 + z^2) (3 t^2 + z^2) \mathit{g}_{\mu \nu }\bigr) \nonumber\\
&& + \tfrac{1}{8} \lambda^4 \bigl(T^{Max}_{\mu \nu } (55 t^4 + 54 t^2 z^2 + 7 z^4) + 4 t (t^2 + z^2) (11 t^2 + 7 z^2) \mathit{g}_{\mu \nu }\bigr) \nonumber\\
&& + \tfrac{1}{32} \lambda^5 \bigl(-2 T^{Max}_{\mu \nu } (273 t^5 + 354 t^3 z^2 + 97 t z^4) - 5 (t^2 + z^2) (91 t^4 + 86 t^2 z^2 + 11 z^4) \mathit{g}_{\mu \nu }\bigr) \nonumber\\
&& + \tfrac{1}{8} \lambda^6 \bigl(T^{Max}_{\mu \nu } (357 t^6 + 575 t^4 z^2 + 243 t^2 z^4 + 17 z^6) + 6 t (t^2 + z^2) (51 t^4 + 64 t^2 z^2 + 17 z^4) \mathit{g}_{\mu \nu }\bigr) \\
&& - \tfrac{1}{64} \lambda^7 \bigl(4 T^{Max}_{\mu \nu } (1938 t^7 + 3729 t^5 z^2 + 2140 t^3 z^4 + 333 t z^6) + 7 (t^2 + z^2) (969 t^6 + 1517 t^4 z^2 + 623 t^2 z^4 + 43 z^6) \mathit{g}_{\mu \nu }\bigr),\nonumber
\end{eqnarray}
where $T^{Max}_{\mu \nu }=F_{\mu }{}^{\alpha } F_{\nu \alpha } - t\, \mathit{g}_{\mu \nu }$. Then we have $ \lambda \frac{\partial{\mathcal{L}_{BN}}}{\partial{\lambda}}=- \frac{1}{4} T^{\mu}{}_{ \mu }$.
To construct an irrelevant operator from the energy-momentum tensor, we use the two structures $T_{\mu \nu } T^{\mu \nu }$ and $T^{\mu}{}_{ \mu } T^{\nu}{}_{ \nu }$, which appear in the Bossard-Nicolai theory as follows.
\begin{eqnarray}\label{TTab}
T_{\mu \nu } T^{\mu \nu } &=& 4 (t^2 + z^2) - 8 t \lambda (t^2 + z^2)+ \lambda^2 (t^2 + z^2) (17 t^2 + 5 z^2) - 8 t \lambda^3 (t^2 + z^2) (5 t^2 + 3 z^2)  \nonumber\\
&& + \lambda^4 (t^2 + z^2) (101 t^4 + 92 t^2 z^2 + 11 z^4) -  \tfrac{1}{2} t \lambda^5 (t^2 + z^2) (535 t^4 + 654 t^2 z^2 + 167 z^4)  \nonumber\\
&&+ \tfrac{1}{8} \lambda^6 (t^2 + z^2) (5865 t^6 + 8999 t^4 z^2 + 3595 t^2 z^4 + 237 z^6)  \nonumber\\
&&-  \tfrac{1}{2} t \lambda^7 (t^2 + z^2) (4123 t^6 + 7613 t^4 z^2 + 4169 t^2 z^4 + 615 z^6)\,,
\end{eqnarray}
and
\begin{eqnarray}\label{TaaTbb}
T^{\mu}{}_{ \mu } T^{\nu}{}_{ \nu } &=& 4 \lambda^2 (t^2 + z^2)^2 - 16 t \lambda^3 (t^2 + z^2)^2  + 4 \lambda^4 (t^2 + z^2)^2 (13 t^2 + 3 z^2)- 80 t \lambda^5 (t^2 + z^2)^2 (2 t^2 + z^2)\nonumber\\
&&  + \tfrac{1}{2} \lambda^6 (t^2 + z^2)^2 (969 t^4 + 762 t^2 z^2 + 73 z^4)- 7 t \lambda^7 (t^2 + z^2)^2 (209 t^4 + 226 t^2 z^2 + 49 z^4)\,.
\end{eqnarray}
By using two structures \ref{TTab} and \ref{TaaTbb}, we can explicitly find the function $f(T_{\mu\nu},\lambda)$ up to order $\lambda^7$ for the Bossard-Nicolai theory in Eq. \ref{BN}  as follows
\begin{equation}
\frac{\partial{\mathcal{L}_{BN}}}{\partial{\lambda}}=\frac{1}{8}\Bigg( T_{\mu \nu } T^{\mu \nu } + \frac{1}{4} T_{\mu }{}^{\mu } T_{\nu }{}^{\nu }-  \frac{1}{16 } \frac{(T_{\mu }{}^{\mu }{} T_{\nu }{}^{\nu }{})^2}{ t_{\mu \nu } t^{\mu \nu }} + \frac{1}{32 }  \frac{(T_{\mu }{}^{\mu }{} T_{\nu }{}^{\nu }{})^3}{ (t_{\mu \nu } t^{\mu \nu })^2}\Bigg),
\end{equation}
where, $t_{\mu \nu }$ is the traceless part of the energy-momentum tensor \ref{TBN}, given by  $t_{\mu \nu }= T_{\mu \nu } -  \frac{1}{4} T^{\alpha }{}_{\alpha } \mathit{g}_{\mu \nu }$.
We can recover the Bossard-Nicolai Lagrangian up to order $\lambda^{12}$ and beyond by choosing $(a_{56}=\tfrac{235927}{768},a_{79}=\tfrac{11025799}{4224})$ in the general Lagrangian \ref{Ap1}.  This power series is derived from the flow equation with the following exact function up to $\lambda^{12}$:
\begin{equation}\label{FBN}
\frac{\partial{\mathcal{L}_{BN}}}{\partial{\lambda}}=\frac{1}{8}\Bigg(T_{\mu \nu } T^{\mu \nu } +\frac{ (t_{\mu \nu } t^{\mu \nu })^2  (T_{\mu }{}^{\mu } T_{\nu }{}^{\nu }) -  \tfrac{1}{4} (t_{\mu \nu } t^{\mu \nu }) ( T_{\mu }{}^{\mu } T_{\nu }{}^{\nu })^2}{4 (t_{\mu \nu } t^{\mu \nu })^2 -\tfrac{1}{2} (T_{\mu }{}^{\mu } T_{\nu }{}^{\nu })^2}\Bigg),
\end{equation} 
the nature of Bossard-Nicolai's theory, which lacks a closed form, necessitates that our findings be presented as a perturbative series.
Consequently, the expansion of Eq. \ref{FBN}  to the order $\lambda^{12} $ aligns with the flow equation derived from Bossard-Nicolai theory.
% \begin{eqnarray}
%T^{\mu}_{ \mu }&=&-2 \lambda (t^2 + z^2) + 4 t \lambda^2 (t^2 + z^2) - 3 \lambda^3 (t^2 + z^2) (3 t^2 + z^2) + 2 t \lambda^4 (t^2 + z^2) (11 t^2 + 7 z^2) \nonumber\\
% &&  -  \tfrac{5}{8} \lambda^5 (t^2 + z^2) (91 t^4 + 86 t^2 z^2 + 11 z^4) + 3 t \lambda^6 (t^2 + z^2) (51 t^4 + 64 t^2 z^2 + 17 z^4)  \nonumber\\
% &&-  \tfrac{7}{16} \lambda^7 (t^2 + z^2) (969 t^6 + 1517 t^4 z^2 + 623 t^2 z^4 + 43 z^6) ,
%\end{eqnarray}
%%%%%%%%%%%%%%%%%%%%%%%%%%%%%%%%%%%%%%%%%%%%%%%%%%%%%%%%%%%%%%%%%%%%%%%%%%%%%%%%%%%%%%%%%%%%%%%%%%%%%%%%%%%%%%%%%%%%%%%%%%%%%%%%%%%%%%%%%%%%%%%%%%%%%%%%%%%%%%%%%%%%%%%%%%%%%%%%%%%%%%%%%%%%%%%%%%%%%%%%%%%%%%%%%%%%%%%%%%%%%%%%%%%%%%%%%%%%%%%%%%%%%%%%%%%%%%%%%%%%%%%%%%%%55
\subsection{Generalizations of irrelevant flow}\label{Secbn}
%%%%%%%%%%%%%%%%%%%%%%%%%%%%%%%%%%%%%%%%%%%%%%%%%%%%%%%%%%%%%%%%%%%%%%%%%%%%%%%%%%%%%%%%%%%%%
In this section, we aim to establish a comprehensive framework for irrelevant $ T\bar{T}$-like deformations that are compatible with the general Lagrangian \ref{Ltz}. Introduced through perturbation, these deformations augment Maxwell's theory with new terms while preserving $ SO(2)$ symmetry.
The flow equation governing all duality-invariant theories within \ref{Ltz}  up to the  order of $\lambda^2$ is given by:
$
 \frac{\partial{{\cal L}(\lambda^n)}}{\partial{\lambda}}=\frac{1}{8}T_{\mu \nu } T^{\mu \nu }+\dots$\,.
This equation is contingent upon the additional terms incorporated into the Lagrangian up to the order of $\lambda^n$. The energy-momentum tensor for \ref{Ltz} can be computed in a general sense, yielding the following result up to $\lambda^7$ order:
\begin{eqnarray}\label{Tmn}
T_{\mu \nu }&=&T^{Max}_{\mu \nu } + \lambda \bigl(- t T^{Max}_{\mu \nu } -  \tfrac{1}{2} (t^2 + z^2) \mathit{g}_ {\mu \nu }\bigr) + \lambda^2 \bigl(\tfrac{1}{2} T^{Max}_{\mu \nu } (3 t^2 + z^2) + t (t^2 + z^2) \mathit{g}_ {\mu \nu }\bigr) \nonumber\\
&& + \lambda^3 \Bigl(- t^3 T^{Max}_{\mu \nu } + \tfrac{3}{4} (- t^4 + z^4) \mathit{g}_{\mu \nu } -  \tfrac{1}{2} a_{11} (t^2 + z^2) \bigl(4 t T^{Max}_{\mu \nu } + 3 (t^2 + z^2) \mathit{g}_{\mu \nu }\bigr)\Bigr)  \nonumber\\
&&+ \lambda^4 \Bigl(2 a_{11} (t^2 + z^2) \bigl(T^{Max}_{\mu \nu } (5 t^2 + z^2) + 4 t (t^2 + z^2) \mathit{g}_{\mu \nu }\bigr) + \tfrac{1}{8} \bigl(- T^{Max}_{\mu \nu } (25 t^4 + 42 t^2 z^2 + 9 z^4) \nonumber\\
&& - 4 t (t^2 + z^2) (5 t^2 + 9 z^2) \mathit{g}_{\mu \nu }\bigr)\Bigr) \nonumber\\
&& + \lambda^5 \Bigl(T^{Max}_{\mu \nu } (4 t^5 - 5 t z^4) + \tfrac{5}{6} (4 t^6 - 15 t^2 z^4 - 11 z^6) \mathit{g}_{\mu \nu } -  \tfrac{1}{3} a_{22} (t^2 + z^2)^2 \bigl(6 t T^{Max}_{\mu \nu } + 5 (t^2 + z^2) \mathit{g}_{\mu \nu }\bigr) \nonumber\\
&& -  \tfrac{5}{3} a_{11} \bigl(6 T^{Max}_{\mu \nu } (t^5 -  t z^4) + 5 (t^6 - 3 t^2 z^4 - 2 z^6) \mathit{g}_{\mu \nu }\bigr)\Bigr) \nonumber\\
&& + \lambda^6 \Bigl(2 a_{11}^2 (t^2 + z^2)^2 \bigl(T^{Max}_{\mu \nu } (7 t^2 + z^2) + 6 t (t^2 + z^2) \mathit{g}_{\mu \nu }\bigr) \nonumber\\
&& + 2 a_{22} (t^2 + z^2)^2 \bigl(T^{Max}_{\mu \nu } (7 t^2 + z^2) + 6 t (t^2 + z^2) \mathit{g}_{\mu \nu }\bigr)  \nonumber\\
&&-  a_{11} (t^2 + z^2) \bigl(T^{Max}_{\mu \nu } (91 t^4 + 154 t^2 z^2 + 23 z^4) + 6 t (t^2 + z^2) (13 t^2 + 23 z^2) \mathit{g}_{\mu \nu }\bigr) \nonumber\\
&& + \tfrac{1}{16} \bigl(T^{Max}_{\mu \nu } (707 t^6 + 1935 t^4 z^2 + 1437 t^2 z^4 + 193 z^6) + 6 t (t^2 + z^2) (101 t^4 + 286 t^2 z^2 + 193 z^4) \mathit{g}_{\mu \nu }\bigr)\Bigr) \nonumber\\
&& + \lambda^7 \Bigl(T^{Max}_{\mu \nu } (-50 t^7 + 189 t^3 z^4 + 140 t z^6) + \tfrac{7}{4} (-25 t^8 + 189 t^4 z^4 + 280 t^2 z^6 + 116 z^8) \mathit{g}_{\mu \nu }\nonumber\\
&& -  \tfrac{1}{4} a_{37} (t^2 + z^2)^3 \bigl(8 t T^{Max}_{\mu \nu } + 7 (t^2 + z^2) \mathit{g}_{\mu \nu }\bigr) \nonumber\\
&& +\tfrac{7}{4} (2 a_{11}^2 + a_{22}) (t^2 + z^2)^2 \bigl(-8 T^{Max}_{\mu \nu } (t^3 - 2 t z^2) - 7 (t^2 - 3 z^2) (t^2 + z^2) \mathit{g}_{\mu \nu }\bigr)  \nonumber\\
&& + \tfrac{7}{2} a_{11} \bigl(4 T^{Max}_{\mu \nu } (8 t^7 - 29 t^3 z^4 - 21 t z^6) + 7 (t^2 + z^2)^2 (4 t^4 - 8 t^2 z^2 - 17 z^4) \mathit{g}_{\mu \nu }\bigr)\Bigr)+{\cal O}(\lambda^8)\,.
\end{eqnarray}
The general trace flow equation, derived from a scalar operator built from the stress tensor within a duality-invariant NED theories, can be obtained from the energy-momentum tensor presented in Eq. \ref{Tmn}. This single trace flow equation is as follows:
\begin{eqnarray}\label{Trmn}
T^{\mu}{}_{ \mu }&=& -4  \lambda  \frac{\partial {\cal L}(\lambda)}{\partial \lambda}\,.
\end{eqnarray}
The trace flow equation in question is universal and independent of the $a_n$ coefficients.
To derive an irrelevant flow equation, we construct two configurations, $T_{\mu \nu } T^{\mu \nu } $ and $ T^{\mu}{}_{ \mu } T^{\nu}{}_{ \nu } $, from the energy-momentum tensor presented in Eq.  \ref{Tmn}. 
The configurations for two structures $T_{\mu \nu } T^{\mu \nu }$ and $T^{\mu}{}_{ \mu } T^{\nu}{}_{ \nu }$ are outlined as follows:
\begin{eqnarray}\label{TTabNGZ}
T_{\mu \nu } T^{\mu \nu } &=&4 (t^2 + z^2) - 8 t \lambda (t^2 + z^2) + \lambda^2 (t^2 + z^2) (17 t^2 + 5 z^2)  \nonumber\\
&&+ \lambda^3 \bigl(-16 a_{11} t (t^2 + z^2)^2 - 8 t (t^2 + z^2) (3 t^2 + z^2)\bigr)\nonumber\\
&&  + \lambda^4 \bigl(2 a_{11} (t^2 + z^2)^2 (51 t^2 + 11 z^2) -  (t^2 + z^2) (t^4 + 32 t^2 z^2 + 11 z^4)\bigr)  \nonumber\\
&& + \lambda^5 \bigl(-12 a_{11} t (19 t^2 -  z^2) (t^2 + z^2)^2 - 16 a_{22} t (t^2 + z^2)^3 + t (t^2 + z^2) (49 t^4 + 66 t^2 z^2 - 7 z^4)\bigr)\nonumber\\
&&  + \lambda^6 \bigl(\tfrac{4}{3} a_{22} (t^2 + z^2)^3 (101 t^2 + 17 z^2) + a_{11}^2 (t^2 + z^2)^3 (137 t^2 + 25 z^2)\nonumber\\
&&  -  \tfrac{1}{3} a_{11} (t^2 + z^2)^2 (1217 t^4 + 3652 t^2 z^2 + 755 z^4) \nonumber\\
&&+ \tfrac{1}{12} (t^2 + z^2) (3083 t^6 + 10005 t^4 z^2 + 8829 t^2 z^4 + 1571 z^6)\bigr) \nonumber\\
&&+ \lambda^7 \bigl(- \tfrac{32}{3} a_{22} t (29 t^2 - 13 z^2) (t^2 + z^2)^3 - 128 a_{11}^2 t (5 t^2 - 2 z^2) (t^2 + z^2)^3 - 16 a_{37} t (t^2 + z^2)^4 \nonumber\\
&& + \tfrac{16}{3} a_{11} t (t^2 + z^2)^2 (319 t^4 + 281 t^2 z^2 - 248 z^4) -  \tfrac{1}{3} t (t^2 + z^2) (2371 t^6 + 4389 t^4 z^2 + 273 t^2 z^4 - 1841 z^6)\bigr)\,,\nonumber\\
&&+{\cal O}(\lambda^8)
\end{eqnarray}
and
\begin{eqnarray}\label{TaaTbbNGZ}
T^{\mu}{}_{ \mu } T^{\nu}{}_{ \nu } &=&4 \lambda^2 (t^2 + z^2)^2 - 16 t \lambda^3 (t^2 + z^2)^2 + \lambda^4 \bigl(4 (7 t^2 - 3 z^2) (t^2 + z^2)^2 + 24 a_{11} (t^2 + z^2)^3\bigr) \nonumber\\
&& + \lambda^5 \bigl(-176 a_{11} t (t^2 + z^2)^3 + 16 t (t^2 + z^2)^2 (t^2 + 6 z^2)\bigr)  \nonumber\\
&&+ \lambda^6 \bigl(\tfrac{4}{3} a_{11} (319 t^2 - 227 z^2) (t^2 + z^2)^3 + 36 a_{11}^2 (t^2 + z^2)^4  \nonumber\\
&&+ \tfrac{80}{3} a_{22} (t^2 + z^2)^4 -  \tfrac{1}{3} (t^2 + z^2)^2 (373 t^4 + 326 t^2 z^2 - 467 z^4)\bigr)  \nonumber\\
&&+ \lambda^7 \bigl(-576 a_{11}^2 t (t^2 + z^2)^4 -  \tfrac{736}{3} a_{22} t (t^2 + z^2)^4  + \tfrac{8}{3} a_{11} t (t^2 + z^2)^3 (341 t^2 + 1181 z^2) \nonumber\\
&&-  \tfrac{2}{3} t (t^2 + z^2)^2 (659 t^4 + 2662 t^2 z^2 + 2339 z^4)\bigr)+{\cal O}(\lambda^8)\,.
\end{eqnarray}
We use the flow equation of the Born-Infeld and Bossard-Nicolai theories as a guide to solving this problem. We add terms of the form $\frac{(T_{\mu }{}^{\mu }{} T_{\nu }{}^{\nu }{})^n}{( t_{\mu \nu } t^{\mu \nu })^{n-1}}$ to the flow equation. This leads us to the following $T \bar{T}$-like deformation for the duality invariant NED theories in Eq. \ref{Ltz}:
\begin{eqnarray}\label{GTTbar}
\frac{\partial {\cal L}(\lambda)}{\partial \lambda}=\tfrac{1}{8} T_{\mu\nu}T^{\mu\nu}+\sum_{n=1} b_n \frac{(T_{\mu }{}^{\mu }{} T_{\nu }{}^{\nu }{})^n}{( t_{\mu \nu } t^{\mu \nu })^{n-1}}\,.
\end{eqnarray}
The types of duality invariant NED theories and the irrelevant form of $T\bar{T}$-like deformation depend on the coefficients $a_m$ and $b_n$. We can relate them by substituting the Lagrangian \ref{Ltz} into the flow equation \ref{GTTbar} and solving for $b_n$ in terms of $a_m$ order by order:
\begin{eqnarray}\label{bn}
&&\bigg(b_1 \to   \frac{3}{8} a_{11}-\frac{11}{32}\bigg) ,
\bigg(b_2 \to 
\frac{1}{48} (77 - 80 a_{11} - 108 a_{11}^2 + 20 a_{22})\bigg),\nonumber\\
&&
\bigg(b_3 \to 
\frac{1}{64} (-1935 + 2590 a_{11} + 1060 a_{11}^2 + 1512 a_{11}^3 - 300 a_{22} - 
480 a_{11} a_{22} + 28 a_{37})\bigg)\,.
\end{eqnarray}
% \begin{eqnarray}\label{GTTbarExpand}
% \frac{\partial {\cal L}(\lambda)}{\partial \lambda}&=&\tfrac{1}{8} T_{\mu\nu}T^{\mu\nu}+( \tfrac{3}{8} a_{11}- \tfrac{11}{32} ) T_{\mu }{}^{\mu }{} T_{\nu }{}^{\nu } + \frac{\bigl(77 - 4 a_{11} (20 + 27 a_{11}) + 20 a_{22}\bigr) }{48 }\frac{(T_{\mu }{}^{\mu }{} T_{\nu }{}^{\nu }{})^2}{( t_{\mu \nu } t^{\mu \nu })}\nonumber\\
% && + \frac{\bigl( 2 a_{11} (1295 + 530 a_{11} + 756 a_{11}^2 - 240 a_{22}) - 300 a_{22} + 28 a_{37}-1935 \bigr) }{64 }\frac{(T_{\mu }{}^{\mu }{} T_{\nu }{}^{\nu }{})^3}{( t_{\mu \nu } t^{\mu \nu })^{2}}
% \end{eqnarray}
Using the coefficients \ref{bn} in the flow equation \ref{GTTbar}, we get a general irrelevant flow equation for all duality invariant electrodynamics theories. We can simplify the problem by substituting $t_{\mu \nu } t^{\mu \nu }= T_{\mu \nu } T^{\mu \nu }-\frac{1}{4} T_{\mu }{}^{\mu }{} T_{\nu }{}^{\nu }$ in \ref{GTTbar} and writing the flow equation \ref{GTTbar} as a power series of $\frac{(T_{\mu }{}^{\mu }{} T_{\nu }{}^{\nu }{})^n}{( T_{\mu \nu } T^{\mu \nu })^{n-1}}$. The general flow equation is then:
\begin{align}\label{GTTbarSeri}
\boxed{ \frac{\partial {\cal L}(\lambda)}{\partial \lambda}=\sum_{n=0}^{\infty} c_n \frac{(T_{\mu }{}^{\mu }{} T_{\nu }{}^{\nu }{})^n}{( T_{\mu \nu } T^{\mu \nu })^{n-1}}\,.}
\end{align}
with coefficients $c_n$ as:
\begin{eqnarray}\label{GTTbarExpandXY}
&&c_0=\tfrac{1}{8},\,\,\,\,\,c_1=( \tfrac{3}{8} a_{11}- \tfrac{11}{32} ),\,\,\,\,\,c_2= \frac{1 }{48 } \bigl(77 - 4 a_{11} (20 + 27 a_{11}) + 20 a_{22}\bigr),\,\,\,\,\nonumber\\
&& c_3=\frac{1 }{96 }\bigl(-2864 + 3845 a_{11} + 1536 a_{11}^2 + 2268 a_{11}^3 - 440 a_{22} - 720 a_{11} a_{22} + 42 a_{37} \bigr)\,.
\end{eqnarray}
Using the method here and the duality invariant Lagrangian in the appendix up to order $\lambda^{12}$, we can obtain the general flow equation in \ref{GTTbarSeri} and all the $c_n$ coefficients as functions of $a_m$ up to order $\lambda^{12}$.
%%%%%%%%%%%%%%%%%%%%%%%%%%%%%%%%%%%%%%%%%%%%%%%%%%%%%%%%%%%%%%%%%%%%%%%%%%%%%%%%%%%%%%%%%%%%%%%%%%%%%%%%%%%%%%%%%%%%%%%%%%%%%%%%%%%%%%%%%%%%%%%%%%%%%%%%%%%%%%%%%%%%%%%%%%%%%%%%%%%%%%%%%%%%%%%%%%%%%%%%%%%%
\section{{Pertubative self-dual solutions with marginal deformations}}\label{Secdn3}

%%%%%%%%%%%%%%%%%%%%%%%%%%%%%%%%%%%%%%%%%%%%%%%%%%%%%%%%%%%%%%%%%%%%%%%%%%%%%%%%%%%%%%%%%%%%%%%%%%%%%%%%%%%%%%%%%%%%%%%%%%%%%%%%%%%%%%%%%%%%%%%%%%%%%%%%%%%%%%%%%%%%%%%%%%%%%%%%%%%%%%%%%%%%%%%%%%%%%
In the previous section, we derived a general Lagrangian (Eq.~\ref{Ltz}) from an irrelevant general deformation (Eq.~\ref{GTTbarSeri}). Now, we explore how to apply this method to marginal deformations. The ModMax Lagrangian represents a marginal deformation of Maxwell's Lagrangian, corresponding to a root $T\bar{T} $-like deformation involving the square root of the energy-momentum tensor. The ModMax theory includes a term of Maxwell's order in the form $\sqrt{z^2 + t^2}$. We investigate a general theory that encompasses all feasible combinations of $ t $, $ z $, and $\sqrt{z^2 + t^2}$ up to the $\lambda^4$ order. This is the Lagrangian we propose as:
\begin{eqnarray}\label{Ltzx}
{\cal L}(\lambda,\gamma )&=&d_{1}{} t  + d_{2}{} \sqrt{z^2 + t^2} + \lambda \bigl(d_{3}{}t^2  +  d_{4}{} t z + d_{5}{} z^2 +d_{6}{} t  \sqrt{z^2 + t^2} + d_{7}{} z \sqrt{z^2 + t^2}\bigr) \nonumber\\
&&+ \lambda^2 \bigl(d_{8}{}t^3  +d_{9}{} t^2  z +d_{10}{} t  z^2 + d_{11}{} z^3 +d_{12}{} t^2  \sqrt{z^2 + t^2} + d_{13}{} z^2 \sqrt{z^2 + t^2}\bigr) \nonumber\\
&&+ \lambda^3 \bigl(d_{14}{}t^4  +d_{15}{} t^3  z + d_{16}{}t^2  z^2 +d_{17}{} t  z^3 + d_{18}{} z^4 + d_{19}{} t^3 \sqrt{z^2 + t^2} \nonumber\\
&& + d_{20}{} t^2 z \sqrt{z^2 + t^2} +d_{21}{} t  z^2 \sqrt{z^2 + t^2} + d_{22}{} z^3 \sqrt{z^2 + t^2}\bigr) \nonumber\\
&&+ \lambda^4 \bigl(d_{23}{} t^5  +d_{24}{}  t^4  z +d_{25}{}  t^3 z^2 + d_{26}{} t^2 z^3 +d_{27}{} t  z^4 + d_{28}{} z^5 +d_{29}{}  t^4 \sqrt{z^2 + t^2} \nonumber\\
&& +d_{30}{} t^3  z \sqrt{z^2 + t^2} +d_{31}{} t^2  z^2 \sqrt{z^2 + t^2} +d_{32}{} t  z^3 \sqrt{z^2 + t^2} + d_{33}{} z^4 \sqrt{z^2 + t^2}\bigr)+{\cal O}(\lambda^5)\,\,.
\end{eqnarray}
By solving the \ref{lagrangeNGZ} differential equations sequentially, we can determine the relation among the $d_n$ coefficients. The requisite conditions for coefficients Lagrangian \ref{Ltzx} to satisfy differential equation \ref{lagrangeNGZ} to the $\lambda^3$ order are outlined below:
\begin{eqnarray}
\label{dn}
&&\Big( d_{2}{} \to \sqrt{ {d_{1}}^2-1}\Big)\,\,\,\,\, ,\,\,\,\,\,\, \Big(d_{3 }\to \frac{\bigl( 2 {d_{1}}^2-1\bigr) d_{5}{}}{{d_{1}}^2}\,\,,\,\,\,d_{6} \to \frac{2  \sqrt{ {d_{1}}^2-1} d_{5}{}}{d_{1}{}}\,\,,\,\,\,d_{4}=d_{7} \to 0\Big) \, ,\nonumber\\
&& \Big(d_{13 }\to -2 {d_{1}}^3 \sqrt{ {d_{1}}^2-1} d_{8}{} -  {d_{1}}^2 d_{12}{} + 2 {d_{1}}^4 d_{12}{}\,\,,\,\,\, 
d_{10 }\to d_{1}{} \Bigl(d_{1}{} \bigl(3 - 2 {d_{1}}^2\bigr) d_{8}{} + 2 \bigl( {d_{1}}^2-1\bigr)^{3/2} d_{12}{}\Bigr)\,\,,\nonumber\\
&& d_{8} \to \frac{2 {d_{5}}^2 -  {d_{1}}^4  \sqrt{ {d_{1}}^2-1} d_{12} + 4 {d_{1}}^6 \sqrt{ {d_{1}}^2-1} d_{12}{}}{{d_{1}}^5 \bigl(4 {d_{1}}^2-3 \bigr)}\,\,,\,\,\,d_{9}=d_{11} \to 0\Big)\nonumber\\
&& \Big(d_{16}\to \frac{4 \bigl(6 - 35 {d_{1}}^2 + 68 {d_{1}}^4 - 72 {d_{1}}^6 + 32 {d_{1}}^8\bigr) {d_{5}}^3}{{d_{1}}^6 \bigl( 28 {d_{1}}^2 - 56 {d_{1}}^4 + 32 {d_{1}}^6-3\bigr)} -  \frac{6 \sqrt{ {d_{1}}^2-1} \bigl(3 - 8 {d_{1}}^2 + 8 {d_{1}}^4\bigr) d_{5}{} d_{12}{}}{{d_{1}}^2 \bigl( 4 {d_{1}}^2-3 \bigr) \bigl(1 - 8 {d_{1}}^2 + 8 {d_{1}}^4\bigr)} + \frac{2 {d_{1}}^2 \bigl( 4 {d_{1}}^2-3 \bigr) d_{14}{}}{1 - 8 {d_{1}}^2 + 8 {d_{1}}^4},
\nonumber\\
&&d_{18} \to \frac{\bigl(4 + 64 {d_{1}}^2 - 64 {d_{1}}^4\bigr) {d_{5}}^3}{3 {d_{1}}^2 - 28 {d_{1}}^4 + 56 {d_{1}}^6 - 32 {d_{1}}^8} + \frac{6 \bigl(1 - 4 {d_{1}}^2\bigr) \sqrt{ {d_{1}}^2-1} d_{5}{} d_{12}{}}{\bigl(4 {d_{1}}^2-3\bigr) \bigl(1 - 8 {d_{1}}^2 + 8 {d_{1}}^4\bigr)} + \frac{{d_{1}}^4 d_{14}{}}{1 - 8 {d_{1}}^2 + 8 {d_{1}}^4} \,\,,\nonumber\\
&& d_{19} \to  \frac{6 d_{5}{} d_{12}{}}{{d_{1}}^3 \bigl( 28 {d_{1}}^2 - 56 {d_{1}}^4 + 32 {d_{1}}^6-3 \bigr)} + \frac{4 \sqrt{ {d_{1}}^2-1} \Bigl(3 \bigl(1 - 4 {d_{1}}^2\bigr) {d_{5}}^3 + {d_{1}}^8 \bigl(3 - 10 {d_{1}}^2 + 8 {d_{1}}^4\bigr) d_{14}{}\Bigr)}{{d_{1}}^7 \bigl( 4 {d_{1}}^2-3\bigr) \bigl(1 - 8 {d_{1}}^2 + 8 {d_{1}}^4\bigr)} ,
\nonumber\\
&&d_{21} \to  \frac{4 \sqrt{ {d_{1}}^2-1} \biggl(\Bigl(8 {d_{1}}^2 \bigl(2 - 7 {d_{1}}^2 + 4 {d_{1}}^4\bigr)-3 \Bigr) {d_{5}}^3 + {d_{1}}^8 \bigl( 4 {d_{1}}^2-3 \bigr) d_{14}{}\biggr)}{{d_{1}}^5 \bigl( 4 {d_{1}}^2-3 \bigr) \bigl(1 - 8 {d_{1}}^2 + 8 {d_{1}}^4\bigr)}- \frac{6 \bigl(3 - 12 {d_{1}}^2 + 8 {d_{1}}^4\bigr) d_{5}{} d_{12}{}}{d_{1}{} \bigl( 28 {d_{1}}^2 - 56 {d_{1}}^4 + 32 {d_{1}}^6-3 \bigr)} ,   \nonumber\\
&&d_{15}=d_{17}=d_{20}=d_{22} \to 0  \Big) \,.
\end{eqnarray}
The required conditions for the order $\lambda^4 $ and the above can be derived using the same method, although they are not presented here due to being excessively crowded. Note that here the coefficients of $d_n$ can be an explicit function of the coupling parameter of $\gamma$. Substituting Eq. \ref{dn} into Lagrangian \ref{Ltzx} and setting $d_{1}=-\cosh(\gamma)$.
Using the change of variable $t=\sinh(\alpha) z$, we can write the general duality-invariant Lagrangian up to order $\lambda^4 $  as:
\begin{eqnarray}\label{Ltzxalpa}
{\cal L}(\lambda,\gamma )&=& - \sinh(\alpha -  \gamma) z + \lambda \bigl(\cosh(\alpha -  \gamma)\bigr)^2 {\sech(\gamma)}^2 d_{5}{} z^2\nonumber\\
&& + \lambda^2 \Bigl(- \frac{2 \bigl(\cosh(\alpha -  \gamma)\bigr)^2 {\sech (\gamma)}^5 \sinh(\alpha + 2 \gamma) {d_{5}}^2 z^3}{ 2 \cosh(2 \gamma)-1} + \frac{\bigl(\cosh(\alpha -  \gamma)\bigr)^3 \sech (\gamma) d_{12}{} z^3}{ 2 \cosh(2 \gamma)-1}\Bigr)\nonumber\\
&& + \lambda^3 \Bigl(\frac{\bigl(\cosh(\alpha -  \gamma)\bigr)^2 \bigl(-3 \cosh(2 \alpha -  \gamma) - 5 \cosh(\gamma) + \cosh(7 \gamma) + 3 \cosh(2 \alpha + 5 \gamma)\bigr) \sech (4 \gamma) {d_{5}}^3 z^4}{\bigl(\cosh(\gamma)\bigr)^7 \bigl( 2 \cosh(2 \gamma)-1\bigr)}\nonumber\\
&& -  \frac{6 \bigl(\cosh(\alpha -  \gamma)\bigr)^3 {\sech (\gamma)}^3 \sech (4 \gamma) \sinh(\alpha + 3 \gamma) d_{5}{} d_{12}{} z^4}{ 2 \cosh(2 \gamma)-1} + \bigl(\cosh(\alpha -  \gamma)\bigr)^4 \sech (4 \gamma) d_{14}{} z^4\Bigr)\nonumber\\
&&+\lambda^4 \biggl(\frac{\bigl(\cosh(\alpha -  \gamma)\bigr)^2{\sech (\gamma)}^{11}  (d_{5}{})^4 z^5}{4 \bigl(1 - 2 \cosh(2 \gamma)\bigr)^2 \cosh(4 \gamma) \bigl(1 - 2 \cosh(2 \gamma) + 2 \cosh(4 \gamma)\bigr)}\nonumber\\
&& \times\Bigl(26 \sinh(\alpha) + 13 \sinh(3 \alpha)- 8 \sinh(\alpha + 14 \gamma)\nonumber\\
&& + \sinh(\alpha - 16 \gamma) - 7 \sinh(\alpha - 10 \gamma) - 23 \sinh(\alpha - 8 \gamma) - 11 \sinh(3 \alpha - 8 \gamma) - 32 \sinh(\alpha - 6 \gamma)\nonumber\\
&& + \sinh(\alpha - 4 \gamma) - 7 \sinh(\alpha - 2 \gamma) + 11 \sinh(3 \alpha - 2 \gamma) + 16 \sinh(\alpha + 2 \gamma) + 11 \sinh\bigl(3 (\alpha + 2 \gamma)\bigr)\nonumber\\
&& + 25 \sinh(\alpha + 4 \gamma) - 11 \sinh\bigl(3 (\alpha + 4 \gamma)\bigr) + 13 \sinh(3 \alpha + 4 \gamma) + 40 \sinh(\alpha + 6 \gamma) + 26 \sinh(\alpha + 8 \gamma) \Bigr)\nonumber\\
&& + \frac{3 \bigl(\cosh(\alpha -  \gamma)\bigr)^3 {\sech (\gamma)}^7 (d_{5}{})^2 d_{12}{} z^5}{4 \bigl(1 - 2 \cosh(2 \gamma)\bigr)^2 \bigl(-1 + 2 \cosh(2 \gamma) - 2 \cosh(4 \gamma)\bigr) \cosh(4 \gamma)}\Bigl(11 \cosh\bigl(2 (\alpha - 2 \gamma)\bigr) + 19 \cosh(2 \gamma) \nonumber\\
&&+ 13 \cosh(4 \gamma) + 3 \cosh(6 \gamma) - 3 \cosh(12 \gamma) + 5 \cosh\bigl(2 (\alpha + \gamma)\bigr) - 5 \cosh\bigl(2 (\alpha + 2 \gamma)\bigr) - 11 \cosh\bigl(2 (\alpha + 5 \gamma)\bigr)\Bigr)\nonumber\\
&& -  \frac{9 \bigl(\cosh(\alpha -  \gamma)\bigr)^4{\sech (\gamma)}^3 \sinh(\alpha + 4 \gamma) (d_{12}{})^2 z^5}{2 \bigl(1 - 2 \cosh(2 \gamma)\bigr)^2 \bigl(1 - 2 \cosh(2 \gamma) + 2 \cosh(4 \gamma)\bigr)} -  \frac{8 \bigl(\cosh(\alpha -  \gamma)\bigr)^4{\sech (\gamma)}^3 \sinh(\alpha + 4 \gamma) d_{5}{} d_{14}{} z^5}{\cosh(4 \gamma) \bigl(1 - 2 \cosh(2 \gamma) + 2 \cosh(4 \gamma)\bigr)}\nonumber\\
&& + \frac{\bigl(\cosh(\alpha -  \gamma)\bigr)^5 \sech (\gamma) d_{29}{} z^5}{1 - 2 \cosh(2 \gamma) + 2 \cosh(4 \gamma)}\biggr)+{\cal O}(\lambda^5)\,.
\end{eqnarray}
 The Lagrangian in Eq.~\ref{Ltzxalpa} reduces to ModMax theory with coupling constant $\gamma$  when $\lambda=0$.
The ModMax theory exhibits a root $ T\bar{T} $-like deformation flow equation. We expect that the duality-invariant Lagrangian in Eq.~\ref{Ltzxalpa} will have a similar flow equation at higher orders of $\lambda$. To study this root deformation flow equation, we need to compute the energy-momentum tensor for the duality-invariant Lagrangian in Eq.~\ref{Ltzxalpa} and construct the $ T_{\mu \nu} T^{\mu \nu} $ and $ T^{\mu}{}_{\mu} T^{\nu}{}_{\nu} $ structures. The explicit form of the  energy-momentum tensor is:
\begin{eqnarray}\label{Tmunuroot}
T_{\mu \nu }(\lambda,\gamma )&=& \cosh(\alpha -  \gamma) \sech (\alpha) T^{Max}_{\mu \nu } \nonumber\\
&& -  \lambda\, d_{5}{} \cosh(\alpha -  \gamma) {\sech (\gamma)}^2  z \bigl(2 \sech (\alpha) \sinh(\alpha -  \gamma) T^{Max}_{\mu \nu } + \cosh(\alpha -  \gamma) z \mathit{g}_{\mu \nu }\bigr) \nonumber\\
&& + \lambda^2 \biggl(- \frac{\bigl(\cosh(\alpha -  \gamma)\bigr)^3 \sech (\gamma) d_{12}{} z^2 }{ 2 \cosh(2 \gamma)-1} \bigl(3 \sech (\alpha) T^{Max}_{\mu \nu } \tanh(\alpha -  \gamma) + 2 z \mathit{g}_{\mu \nu }\bigr) \nonumber\\
&&+ \frac{\cosh(\alpha -  \gamma) {\sech (\gamma)}^5 {d_{5}}^2 z^2 }{ 2 \cosh(2 \gamma)-1} \Bigl(\bigl(- \cosh(3 \gamma) + 3 \cosh(2 \alpha + \gamma)\bigr) \sech (\alpha) T^{Max}_{\mu \nu } \nonumber\\
&&+ 2 \bigl(\sinh(3 \gamma) + \sinh(2 \alpha + \gamma)\bigr) z \mathit{g}_{\mu \nu }\Bigr)\biggr)\nonumber\\
&& + \lambda^3 \Biggl(- \bigl(\cosh(\alpha -  \gamma)\bigr)^3 \sech (4 \gamma) d_{14}{} z^3 \bigl(4 \sech (\alpha) \sinh(\alpha -  \gamma) T^{Max}_{\mu \nu } + 3 \cosh(\alpha -  \gamma) z \mathit{g}_{\mu \nu }\bigr)\nonumber\\
&& + \frac{\cosh(\alpha -  \gamma) {\sech (\gamma)}^7 \sech (4 \gamma) {d_{5}}^3 z^3 }{-2 + 4 \cosh(2 \gamma)} \Bigl(-2 \sech (\alpha) \bigl(-5 \sinh(\alpha) + \sinh(\alpha - 8 \gamma)\nonumber\\
&&  - 5 \sinh(\alpha - 2 \gamma)- 6 \sinh(3 \alpha - 2 \gamma) + 6 \sinh(3 \alpha + 4 \gamma) + \sinh(\alpha + 6 \gamma)\bigr) T^{Max}_{\mu \nu } - 3 \bigl(-8 \cosh(\alpha) \nonumber\\
&&+ \cosh(\alpha - 8 \gamma) - 5 \cosh(\alpha - 2 \gamma) - 3 \cosh(3 \alpha - 2 \gamma) + 3 \cosh(3 \alpha + 4 \gamma) + 4 \cosh(\alpha + 6 \gamma)\bigr) z \mathit{g}_{\mu \nu }\Bigr)\nonumber\\
&&+ \frac{3 \bigl(\cosh(\alpha -  \gamma)\bigr)^2 \sech (\alpha) {\sech (\gamma)}^3 d_{5}{} d_{12}{} z^3 }{-2 + 4 \cosh(2 \gamma)} \biggl(\Bigl(-4 + 8 \cosh\bigl(2 (\alpha + \gamma)\bigr) \sech (4 \gamma)\Bigr) T^{Max}_{\mu \nu }\nonumber\\
&& + 3 \sech (4 \gamma) \bigl(\sinh(\alpha + 2 \gamma) - \sinh(\alpha - 4 \gamma) + \sinh(3 \alpha + 2 \gamma) + \sinh(\alpha + 4 \gamma)\bigr) z \mathit{g}_{\mu \nu }\biggr)\Biggr)\nonumber\\
&&+{\cal O}(\lambda^4),
\end{eqnarray}
Where $O(\lambda^4)$ are $\lambda^4$ orders of energy-momentum tensors, which are shown in the appendix in Eq.\ref{O4}.
From Eq. \ref{Tmunuroot}, we obtain the following expressions for the  $T_{\mu \nu } T^{\mu \nu }$ and $T^{\mu}{}_{ \mu } T^{\nu}{}_{ \nu }$  structures:
\begin{eqnarray}\label{TTabMarg}
	T_{\mu \nu } T^{\mu \nu } &=&4 \bigl(\cosh(\alpha -  \gamma)\bigr)^2 z^2 - 16 \lambda \bigl(\cosh(\alpha -  \gamma)\bigr)^2 {\sech (\gamma)}^2 \sinh(\alpha -  \gamma) d_{5}{} z^3  \nonumber\\
	&& + \lambda^2 \Bigl(\frac{\bigl(\cosh(\alpha -  \gamma)\bigr)^2 {\sech (\gamma)}^5 {d_{5}}^2 z^4}{ 2 \cosh(2 \gamma)-1}  \bigl(5 \cosh(2 \alpha - 5 \gamma) - 14 \cosh(3 \gamma) + 29 \cosh(2 \alpha + \gamma)\bigr)\nonumber\\
	&& - 24 \bigl(\cosh(\alpha -  \gamma)\bigr)^3 \sech (3 \gamma) \sinh(\alpha -  \gamma) d_{12}{} z^4\Bigr) \nonumber\\
	&& + \lambda^3 \Biggl(- \frac{16 \bigl(\cosh(\alpha -  \gamma)\bigr)^2 {\sech (\gamma)}^7 \sech (4 \gamma)  {d_{5}}^3 z^5}{ 2 \cosh(2 \gamma)-1} \nonumber\\
	&&  \times \Bigl(-3 \sinh(\alpha) + \sinh(3 \alpha - 4 \gamma) - 6 \cosh(\alpha) \sinh\bigl(2 (\alpha -  \gamma)\bigr) + 4 \sinh(3 \alpha + 4 \gamma)\Bigr)\nonumber\\
	&&+ \frac{16 \bigl(\cosh(\alpha -  \gamma)\bigr)^3 {\sech (\gamma)}^3  d_{5}{} d_{12}{} z^5}{ 2 \cosh(2 \gamma)-1}\biggl( \Bigl(\cosh\bigl(2 (\alpha - 3 \gamma)\bigr) + 7 \cosh\bigl(2 (\alpha + \gamma)\bigr)\Bigr) \sech (4 \gamma)-4 \biggr)\nonumber\\
	&& - 32 \bigl(\cosh(\alpha -  \gamma)\bigr)^4 \sech (4 \gamma) \sinh(\alpha -  \gamma) d_{14}{} z^5\Biggr)+{\cal O}(\lambda^4),
\end{eqnarray}
and
\begin{eqnarray}\label{Ttmumu}
	T^{\mu}{}_{ \mu } T^{\nu}{}_{ \nu }&=& 16 \lambda^2 \bigl(\cosh(\alpha -  \gamma)\bigr)^4 {\sech (\gamma)}^4 {d_{5}}^2 z^4 \nonumber\\
	&& + \lambda^3 \Bigl(\frac{64 \bigl(\cosh(\alpha -  \gamma)\bigr)^5 {\sech (\gamma)}^3 d_{5}{} d_{12}{} z^5}{ 2 \cosh(2 \gamma)-1}- \frac{128 \bigl(\cosh(\alpha -  \gamma)\bigr)^4 {\sech (\gamma)}^7 \sinh(\alpha + 2 \gamma) {d_{5}}^3 z^5}{ 2 \cosh(2 \gamma)-1} \Bigr)\nonumber\\
    &&
    +{\cal O}(\lambda^4)\,.
\end{eqnarray}
 The duality-invariant Lagrangian \ref{Ltzxalpa} is a generic case of theories that obey the root flow equation.
We seek the general form of root deformations and the $d_n$ coefficients that allow it.
\begin{align}\label{GRoot}
\boxed{\frac{\partial {\cal L}(\lambda,\gamma )}{\partial \gamma}= \frac{1}{2}
\sqrt{ \sum_{n=0}^{\infty} e_n \frac{(T_{\mu }{}^{\mu }{} T_{\nu }{}^{\nu }{})^n}{( T_{\mu \nu } T^{\mu \nu })^{n-1}}}\,.}
\end{align}
A relation between the coefficients $d_n$ and $ e_n $ is feasible. The coefficients $ d_n(\gamma) $ depend on $\gamma$, so both $ d_n(\gamma) $ and its derivative ${d}'_{n}(\gamma) = \frac{\partial d_n(\gamma)}{\partial \gamma} $ influence the left side of Eq.~\ref{GRoot}. In contrast, the right side of Eq.~\ref{GRoot} depends only on $d_n(\gamma)$ and not on ${d}'_{n}(\gamma) $. To determine the relationship between the $ d_n(\gamma) $ and $ e_n $ coefficients, we need to solve a first-order differential equation for each order of $\lambda$. This approach allows us to express all the $ d_n(\gamma) $ coefficients in terms of the $ e_n $ coefficients. The Lagrangian ${\cal L}(\lambda, \gamma)$ in Eq.~\ref{GRoot} contains no $ d_n $ coefficients at the order of $\lambda^0$, and it is straightforward to show that $ e_0 = 1 $.
The first order of $\lambda$ differential equation gives us $d_5$ in the next order of Eq. \ref{GRoot}. The differential equation is:
\begin{eqnarray}\label{d5eq}
2\, d_5 \tanh(\gamma) -  {d}'_{5}=0\,.
\end{eqnarray}
 The solution of Eq. \ref{d5eq} is:
\begin{eqnarray}\label{d5}
d_5= n_{1} {\cosh(\gamma)}^2\,.
\end{eqnarray}
The differential equation \ref{d5eq} gives $d_5$ as a function of $\gamma$ with a constant $n_1$. The structure  $T_{\mu }{}^{\mu }{} T_{\nu }{}^{\nu }{}$  shares orders of $\lambda^2$ in Eq.  \ref{GRoot}, so $e_1$ is absent in $d_5$.
The general root flow equation \ref{GRoot} with $d_5$ from \ref{d5} produces an $\lambda^2$th order differential equation:
\begin{eqnarray}\label{d12eq}
 12 d_{12} \sinh(3 \gamma) +{n_{1}}^2 \bigl(25 + 4 e_{1}{} + (1 + 4 e_{1}{})  \cosh(6 \gamma)\bigr)  - 4 \cosh(3 \gamma){d}'_{12}=0\,.
\end{eqnarray}
The differential equation \ref{d12eq} gives $d_{12}$ as a function of $\gamma$, $n_1$, $n_2$, and $e_1$:
\begin{eqnarray}\label{d12}
d_{12}=n_{2}{} \cosh(3 \gamma)  + \tfrac{1}{4}  {n_{1}}^2  \cosh(3 \gamma) \bigl(2 \gamma (1 + 4 e_{1}{}) + 8 \tanh(3 \gamma)\bigr)\,,
\end{eqnarray}
where  $n_1$, $n_2$, and $e_1$ are constants.
%\begin{eqnarray}\label{d14eq}
%&&8 d_{14} \sinh(4 \gamma) + 3  (1 + 4 e_{1}{}) {n_{1}}^3 \sinh(8 \gamma) - 2 \cosh(4 \gamma) %{d}'_{14}\nonumber\\
%&& +n_{1}{}  \bigl(25 + 4 e_{1}{} + \cosh(8 \gamma) (1 + 4 e_{1}{})\bigr) \times \bigl(\gamma (1 + 4 e_{1}{}) {n_{1}}^2 + 2 n_{2}{}\bigr)=0\,.
%\end{eqnarray}
This method finds all the $d_n$ coefficients in higher orders. The $d_{14}$ and $d_{29}$ coefficients are:
\begin{eqnarray}\label{d14}
d_{14}=\tfrac{1}{2} \cosh(4 \gamma) \Bigl((\gamma + 4 \gamma e_{1}{})^2 {n_{1}}^3 + 4 \gamma (1 + 4 e_{1}{}) n_{1}{} n_{2}{} + 2 n_{3}{} + 6 n_{1}{} \bigl(\gamma (1 + 4 e_{1}{}) {n_{1}}^2 + 2 n_{2}{}\bigr) \tanh(4 \gamma)\Bigr)\,,
\end{eqnarray}
and
%\begin{eqnarray}\label{d29eq}
%&& 80 d_{29}  \sinh(5 \gamma) + \bigl(-2089 - 40 e_{1}{} - 16 {e_{1}}^2 + 2 (213 + 32 e_{1}{}) (\gamma + 4 \gamma e_{1}{})^2 + 64 e_{2}{}\bigr) {n_{1}}^4+ 8 (83 + 12 e_{1}{}) n_{1}{} n_{3}{} \nonumber\\
%&& + 8 \gamma (1 + 4 e_{1}{}) (213 + 32 e_{1}{}) {n_{1}}^2 n_{2}{} + 8 (47 + 8 e_{1}{}) {n_{2}}^2 + 82 \sinh(10 \gamma) (1 + 4 e_{1}{}) {n_{1}}^2 \bigl(\gamma (1 + 4 e_{1}{}) {n_{1}}^2 + 2 n_{2}{}\bigr)\nonumber\\
%&&+ \cosh(10 \gamma) \Bigl(\bigl(-9 - 40 e_{1}{} - 16 {e_{1}}^2 + 16 \gamma^2 (1 + 4 e_{1}{})^3 + 64 e_{2}{}\bigr) {n_{1}}^4 + 64 \gamma (1 + 4 e_{1}{})^2 {n_{1}}^2 n_{2}{}\nonumber\\
%&& + 16 (1 + 4 e_{1}{}) {n_{2}}^2 + 24 (1 + 4 e_{1}{}) n_{1}{} n_{3}{}\Bigr) - 16 \cosh(5 \gamma) {d}'_{29}=0
%\end{eqnarray}
\begin{eqnarray}\label{d29}
 d_{29}&=& \tfrac{3}{24} \sinh(5 \gamma) \Bigl(\bigl(-208 + 41 (\gamma + 4 \gamma e_{1}{})^2\bigr) {n_{1}}^4 + 164 \gamma (1 + 4 e_{1}{}) {n_{1}}^2 n_{2}{} + 36 {n_{2}}^2 + 64 n_{1}{} n_{3}{}\Bigr)\nonumber\\
&& +\tfrac{\gamma}{24}\, \cosh(5 \gamma)  \biggl(\Bigl(16 \gamma^2 (1 + 4 e_{1}{})^3 - 3 \bigl(9 + 8 e_{1}{} (5 + 2 e_{1}{}) - 64 e_{2}{}\bigr)\Bigr) {n_{1}}^4\nonumber\\
&& + 96 \gamma (1 + 4 e_{1}{})^2 {n_{1}}^2 n_{2}{} + 48 (1 + 4 e_{1}{}) {n_{2}}^2 + 72 (1 + 4 e_{1}{}) n_{1}{} n_{3}{}\biggr) +  n_{4}{}  \cosh(5 \gamma)\,.
\end{eqnarray}
We consider a general duality-invariant non-linear electrodynamic theory described by the Lagrangian \ref{Ltzxalpa}, where the coefficients $d_5$, $d_{12}$, $d_{14}$ and $d_{29}$ are functions of $\gamma$ given by Eqs. \ref{d5}, \ref{d12},\ref{d14} and \ref{d29}, respectively. The general root flow equation of this duality-invariant theory is given by \ref{GRoot}, where the coefficients $e_n$ can be verified by direct calculation.
%%%%%%%%%%%%%%%%%%%%%%%%%%%%%%%%%%%%%%%%%%%%%%%%%%%%%%%%%%%%%%%%%%%%%%%%%%%%%%%%%%
\subsection{New root  flow equation with negative sign respect to $\alpha$}
%%%%%%%%%%%%%%%%%%%%%%%%%%%%%%%%%%%%%%%%%%%%%%%%%%%%%%%%%%%%%%%%%%%%%%%%%%%%%%%%%%%%%%%%%%%%%%
The change of variable $t=\sinh(\alpha) z$, 
introduces the $ \alpha$ parameter as a pseudo-coupling parameter in the Lagrangian. Investigating the flow equation for this pseudo-coupling parameter presents an intriguing prospect. Given that the $ \alpha$ parameter, akin to $\gamma$, manifests as an angle within the Lagrangian, it is anticipated that the flow equation will bear a resemblance to that of $\gamma$. A particular selection is made to derive this flow equation, detailed below:
\begin{eqnarray}\label{en1}
e_0=1 \,\,\,\,\,\,e_1=-\tfrac{1}{4}\,\,\,\,\,\, e_2=e_3=\dots=0 \,.
\end{eqnarray}
Upon selecting the \ref{en1}, the $d_n$ coefficients are as follows:
\begin{eqnarray}\label{dnen}
&&d_{5} = n_1 \, {\cosh(\gamma)}^2\,,~~~ d_{12}(\gamma)=2 n^2_{1} \sinh(3 \gamma)  +n_{2}  \cosh(3 \gamma)\, ,~~~ d_{14}=6 n_{1}{} n_{2}{} \sinh(4 \gamma) +n_{3}{} \cosh(4 \gamma) \,, \nonumber\\
&&d_{29}=\tfrac{1}{2}  \bigl(-52 n^4_{1} + 9 n^2_{2} + 16 n_{1} n_{3}{}\bigr) \sinh(5 \gamma) + n_{4}{} \cosh(5 \gamma)\,. 
\end{eqnarray}
Substituting the coefficients $d_n$ from Eq. \ref{dnen} into the general Lagrangian \ref{Ltzxalpa} and Eqs. \ref{TTabMarg} and \ref{Ttmumu}, we obtain a novel root flow equation that exhibits a negative sign with respect  to $\alpha$ as:
\begin{align}\label{GTTalpa}
\boxed{ \frac{\partial {\cal L}(\lambda,\gamma )}{\partial \alpha}=- \frac{1}{2}
	\sqrt{T_{\mu\nu}T^{\mu\nu}- \frac{1}{4} {T_{\mu}}^{\mu} {T_{\nu}}^{\nu}}\,.}
\end{align}
This flow equation is identical to the flow equation with respect to $\gamma$, distinguished only by a negative sign; that is, $ \frac{\partial {\cal L}(\lambda,\gamma )}{\partial \alpha}=-\frac{\partial {\cal L}(\lambda,\gamma )}{\partial \gamma} $.
%%%%%%%%%%%%%%%%%%%%%%%%%%%%%%%%%%%%%%%%%%%%%%%%%%%%%%%%%%%%%%%%%%%%%
\subsection{Commutative flows}
%%%%%%%%%%%%%%%%%%%%%%%%%%%%%%%%%%%%%%%%%%%%%%%%%%%%%%%%%%%%%%%%%%
The NED theories have two kinds of flow equations: general irrelevant and general marginal. We derived these two marginal and irrelevant $T\bar{T}$-like deformations for theories that preserve duality symmetry.
We consider the general duality invariant theory in Eq. \ref{Ltzxalpa} with $d_n$ coefficients in Eqs. \ref{d5}, \ref{d12},\ref{d14} and \ref{d29}, which has a general root flow equation with constants $e_n$ and $n_n$. We want this Lagrangian to also have an irrelevant flow equation like \ref{GTTbarSeri}, besides the marginal one. We use $d_n$ to write a general equation of irrelevant flow for the Lagrangian \ref{Ltzxalpa}, which we derived in Eqs. \ref{d5}, \ref{d12},\ref{d14} and \ref{d29}. Solving this equation order by order, we find the values of $e_n$ and $n_n$ as:
\begin{eqnarray}\label{encn1}
&&e_0=1, \,\,\,\,\,\,e_1=-\tfrac{1}{4}, \,\,\,\,\,\, e_2=e_3=\dots=0, \nonumber\\
&&n_1=\tfrac{1}{2}, \,\,\,\,\,n_2=0, \,\,\,\,\,n_3=\tfrac{1}{4} (2 a_{11}+1),\,\,\,\,\,n_4=0.
\end{eqnarray}
Using Eq. \ref{encn1} in Eqs.  \ref{d5}, \ref{d12},\ref{d14} and \ref{d29}, we obtain the values of $d_n$:
\begin{eqnarray}\label{dne1}
&&d_{5}(\gamma) = \tfrac{1}{2}\, {\cosh(\gamma)}^2, \,\,\,\,\,\,\,\,\,\,\,\,\,\,\,\,\,\,\,\,\,\,\,\,\,\,\,\,\,\,\,\,\,\,\,\,\,\,\,\,\,d_{12}(\gamma)=\frac{1}{2} \,\sinh(3 \gamma),\nonumber\\
&&d_{14}(\gamma)=\tfrac{1}{4}(2 a_{11}+1) \, \cosh(4\gamma),\,\,\,\,\,\,\,\,\,\,\,\,d_{29}=( 2 a_{11}- \tfrac{5}{8} ) \sinh(5 \gamma)\,. 
\end{eqnarray}
Upon substituting coefficient \ref{dne1} into the general Lagrangian \ref{Ltzxalpa} and reformulating it in ModMax terms, we obtain a duality-invariant Lagrangian, presented here: 
\begin{eqnarray}\label{LtzGBI1}
	{\cal L}(\lambda,\gamma )&=& {\cal L}_{MM} + \tfrac{1}{2} \lambda \bigl({\cal L}_{MM}^2 + z^2\bigr) + \tfrac{1}{2} \lambda^2 {\cal L}_{MM} \bigl({\cal L}_{MM}^2 + z^2\bigr)\nonumber\\
	&&+ \lambda^3  \bigl(\tfrac{1}{2} a_{11} ({\cal L}_{MM}^2 + z^2)^2 + \tfrac{1}{4} ({\cal L}_{MM}^4 -  z^4)\bigr)\nonumber\\
	&& + \lambda^4 \bigl(2 a_{11} {\cal L}_{MM} ({\cal L}_{MM}^2 + z^2)^2 - \tfrac{1}{8} {\cal L}_{MM} ({\cal L}_{MM}^2 + z^2) (5 {\cal L}_{MM}^2 + 9 z^2)\bigr)+O(\lambda^5).
\end{eqnarray}
Lagrangian \ref{LtzGBI1} becomes Lagrangian \ref{Ltz} when $\gamma \to 0$. The generalized Lagrangian, featuring two independent flow equations (irrelevant and marginal) corresponding to coupling parameters $\lambda$ and $\gamma$, is derived by substituting $ (-t )\rightarrow ({\cal L}_{MM} ) $ in the overarching Lagrangian of \ref{Ap1}, valid through the order of $ \lambda^{12} $. For the specified Lagrangian, the general flow equation is as follows:
\begin{align}\label{GTTbarSerim}
\boxed{ \frac{\partial {\cal L}(\lambda,\gamma)}{\partial \lambda}=\sum_{n=0}^{\infty} c_n \frac{(T_{\mu }{}^{\mu }{} T_{\nu }{}^{\nu }{})^n}{( T_{\mu \nu } T^{\mu \nu })^{n-1}}\,:\,\,\,\,\,\,\,\,\,\,\,\,\,\,\,\,\,\,\frac{\partial {\cal L}(\lambda,\gamma )}{\partial \gamma}= \frac{1}{2}
	\sqrt{T_{\mu\nu}T^{\mu\nu}- \frac{1}{4} {T_{\mu}}^{\mu} {T_{\nu}}^{\nu}}\,,}
\end{align}
with coefficients $c_n$ as:
\begin{eqnarray}\label{GTTbarExpandXmY}
&&c_0=\tfrac{1}{8},\,\,\,\,\,c_1=( \tfrac{3}{8} a_{11}- \tfrac{11}{32} ),\,\,\,\,\,c_2= \frac{1 }{48 } \bigl(77 - 4 a_{11} (20 + 27 a_{11}) + 20 a_{22}\bigr),\,\,\,\,\nonumber\\
&& c_3=\frac{1 }{96 }\bigl(-2864 + 3845 a_{11} + 1536 a_{11}^2 + 2268 a_{11}^3 - 440 a_{22} - 720 a_{11} a_{22} + 42 a_{37} \bigr).
\end{eqnarray}
%\subsection{Some non-trivial flow equations}
Utilizing the general flow equation \ref{GTTbarSerim}, we can formulate comprehensive flow equations that remain applicable across varying coefficients $d_n$ to any specified order of $ \lambda$. Our focus is on deriving certain non-trivial flow equations up to the  $\lambda^{12}$ order. The flow equations are structured to commute with the root operator, ensuring that the root current equation pertaining to coupling $\gamma$  remains autonomous. The following are examples of the aforementioned flow equations:
\begin{itemize}

\item To streamline our discussion, we define two variables: $X = T_{\mu \nu} T^{\mu \nu}$ and $Y = T^{\mu}{}_{\mu} T^{\nu}{}_{\nu}$. In the context of the Lagrangian referenced in \ref{Ap1}, we modify it by substituting $(-t )\rightarrow ({\cal L}_{MM} ) $. This allows us to select the coefficients $ a_n $ in table \ref{T11} to yield a distinct Lagrangian flow equation that mirrors the series expansion of an exponential function. The flow equation, characterized by its exponential nature up to $\lambda^{12}$, is presented below:
\begin{eqnarray}\label{non0}
\frac{\partial {\cal L}(\lambda,\gamma)}{\partial \lambda}&=&\frac{1}{8} e^{-\tfrac{2 Y}{X}} X=\tfrac{1}{8} X -  \tfrac{1}{4} Y + \frac{Y^2}{4 X} -  \frac{Y^3}{6 X^2} + \frac{Y^4}{12 X^3} -  \frac{Y^5}{30 X^4}+\dots,\\
&&c_0=\tfrac{1}{8},\,\,\,\,\,c_1=-\frac{1}{4} ,\,\,\,\,\,c_2= \frac{1 }{4 },\,\,\dots\,.\nonumber
\end{eqnarray} 
This exponential flow equation commutes with the root flow equation presented in Eq. \ref{GTTbarSerim}.

\item An alternative category of nonlinear flow equations manifests in fractional form. The expansion of this fractional flow equation unfolds as follows:
\begin{eqnarray}\label{non1}
\frac{\partial {\cal L}(\lambda,\gamma)}{\partial \lambda}&=&\frac{X^2}{8 X -  Y}=\tfrac{1}{8} X + \tfrac{1}{64} Y + \frac{Y^2}{512 X} + \frac{Y^3}{4096 X^2} + \frac{Y^4}{32768 X^3} + \frac{Y^5}{262144 X^4}+\dots\,,\\
&&c_0=\tfrac{1}{8},\,\,\,\,\,c_1=\frac{1}{64} ,\,\,\,\,\,c_2= \frac{1 }{512 },\,\,\dots\,,\nonumber
\end{eqnarray} 
where, the coefficients of $c_n$ have the general form of $c_n \sim \frac{1}{8} \frac{1}{2^{3n}}$.
\item By selecting coefficients of $a_n$ in table \ref{T11}, we identify a set of polylogarithmic flow equations. From Lagrangian \ref{Ap1} up to $\lambda^{12}$, we derive a theory encompassing the subsequent non-trivial flow equation:  
\begin{eqnarray}\label{non2}
\frac{\partial {\cal L}(\lambda,\gamma)}{\partial \lambda}&=&- \frac{X^2 }{8 Y} PoLog(2, - \frac{Y}{X})=\tfrac{1}{8} X -  \tfrac{1}{32} Y + \frac{Y^2}{72 X} -  \frac{Y^3}{128 X^2} + \frac{Y^4}{200 X^3} -  \frac{Y^5}{288 X^4}+\dots,\\
&&c_0=\tfrac{1}{8},\,\,\,\,\,c_1=-\frac{1}{32} ,\,\,\,\,\,c_2= \frac{1 }{72 },\dots,\nonumber
\end{eqnarray} 
\item Finally, we formulate a theory with a logarithmic flow equation. The development of this flow equation is delineated below: 
\begin{eqnarray}\label{non}
\frac{\partial {\cal L}(\lambda,\gamma)}{\partial \lambda}&=&- \frac{X^2}{8 Y}  \log(\frac{X -  Y}{X})=\tfrac{1}{8} X + \tfrac{1}{16} Y + \frac{Y^2}{24 X} + \frac{Y^3}{32 X^2} + \frac{Y^4}{40 X^3} + \frac{Y^5}{48 X^4}+\dots,\\
&&c_0=\tfrac{1}{8},\,\,\,\,\,c_1=\frac{1}{16} ,\,\,\,\,\,c_2= \frac{1 }{24 },\dots\,,\nonumber
\end{eqnarray} 
\end{itemize}
\begin{table}[h!]
	\begin{center}
		\renewcommand*{\arraystretch}{1.5}
		\begin{tabular}{|c|c|c|c|c|c|}\hline
			Flow Equation & $a_{11}$  &$a_{22}$  &$a_{37}$  &$a_{56}$  &$a_{79}$ \\ \hline
			$\frac{\partial {\cal L}(\lambda,\gamma)}{\partial \lambda}=\tfrac{1}{8} e^{-\frac{2 Y}{X}} X$ &  $\frac{1}{4}$ & $-\frac{153}{80} $ &$\frac{9113}{672}$  &$-\frac{3599567}{80640}$  &$-\frac{2387529037}{8870400}$ \\  \hline
			$\frac{\partial {\cal L}(\lambda,\gamma)}{\partial \lambda}=\frac{X^2}{8 X -  Y}$ & $\frac{23}{24}$  &$ \frac{9499}{1920}$ &$\frac{261691}{8064}$  &$\frac{1114881517}{4644864}$  &$\frac{1393382483561}{729907200}$ \\ \hline
			$\frac{\partial {\cal L}(\lambda,\gamma)}{\partial \lambda}=- \frac{X^2 }{8 Y} PoLog(2, - \frac{Y}{X})$ & $\frac{5}{6}$ &$\frac{49}{15}$  &$\frac{129}{8}$  &$\frac{20812883}{233280}$  &$\frac{27119488703}{51321600}$\\ \hline
			$\frac{\partial {\cal L}(\lambda,\gamma)}{\partial \lambda}=- \frac{X^2}{8 Y}  \log(\frac{X -  Y}{X})$ & $\frac{13}{12}$  & $\frac{1661}{240}$  &$\frac{118021}{2016}$  &$\frac{137316619}{241920}$  &$\frac{158855074919}{26611200}$ \\ \hline
		\end{tabular}
		\caption{ \label{T11} Some non-trivial irrelevant flow equations in  general duality invariant NED theories. The $X=T_{\mu \nu } T^{\mu \nu }$ and $Y=T^{\mu}{}_{ \mu } T^{\nu}{}_{ \nu }$, are considered. }
	\end{center}
\end{table}

\section{Conclusions and perspects}\label{04}
In this paper, we have developed a comprehensive framework for understanding $T \overline{T}$-like deformations in duality-invariant  NED theories. By employing a high-order perturbation approach, we have classified two primary types of solutions that satisfy the differential self-duality condition. Our findings include both irrelevant stress tensor flows, similar to $T \overline{T}$ deformations, and a combination of irrelevant and marginal root-$T \overline{T}$-like deformations. These results significantly extend our understanding of the role of duality invariance in NED theories and offer new perspectives on how these theories can be deformed while preserving crucial physical properties. Our analysis of the Bossard-Nicolai theory and the ModMax theory further demonstrates the applicability and robustness of our approach. The identification of new non-trivial flow equations within these frameworks illustrates the potential for future research to explore more complex and rich structures in nonlinear electrodynamics.

\iffalse
The main points obtained in the current work are the following:

\begin{enumerate}
    \item \textbf{Development of General Framework:} We have formulated a general framework using the Gaillard-Zumino approach to derive duality-invariant NED theories that can accommodate $T \overline{T}$-like deformations.
    
    \item \textbf{Classification of Solutions:} Through high-order perturbation methods, we classified solutions that satisfy the differential self-duality condition, identifying irrelevant stress tensor flows and combinations of irrelevant and marginal root-$T \overline{T}$-like deformations.
    
    \item \textbf{Generalized Flow Equations:} We derived generalized flow equations for duality-invariant NED theories, providing deeper insights into irrelevant and marginal deformations.
    
    \item \textbf{Analysis of Bossard-Nicolai Theory:} We identified a non-trivial flow equation within the Bossard-Nicolai theory, showcasing the versatility of our framework.
    
    \item \textbf{Examination of ModMax Theory:} We explored the ModMax theory, extending Maxwell's electrodynamics into the nonlinear regime while preserving SO(2) and conformal invariance, and provided detailed perturbative expansions and flow equations relevant to this theory.
    
\end{enumerate}
\fi

Recent studies, as cited in  Refs. \cite{Conti:2022egv, Morone:2024egv, Babaei2024c}, have shown that modified gravity models can emerge from $T\bar{T}$-like deformations. These works have delved into the coupling of gravity and the metric implications of quadratic $T\bar{T}$-like perturbations, with a keen focus on their relevance to modified gravity models. The generalized form of the $T\bar{T} $-like deformation, as derived in Eq. \ref{GTTbarSerim}, serves as a compelling impetus to investigate novel gravitational theories stemming from such deformations. Our findings suggest that a similar interpretative approach is applicable. Advancing this research, it would be advantageous to develop a perturbative metric expansion and ascertain if gravitational dressing can be analytically performed, paralleling the methodologies explored in Refs. \cite{Morone:2024egv,Babaei2024c}.

The study in \cite{Conti:2022egv} presents an intermediate scenario where NED theories are reduced from four dimensions to two. This reduction process yields a theory identical to one derived from transforming a group of free scalars via a two-dimensional root-$T\bar{T} $ flow, as detailed in \cite{Babaei-Aghbolagh:2022kjj,Ferko:2206jsw}. The resulting ``{\it modified scalar}''  theory will be the focus of our future research.

A pivotal inquiry addressed whether alternatives to the Born-Infeld theory exist that satisfy the self-duality condition and possess an explicit closed form. Recent findings by Russo and Townsend in \cite{russo2024causal,russo2024dual} have resolved this, revealing multiple causal self-dual electrodynamics theories with closed-form expressions. This revelation prompts further questions regarding the flow equations governing these theories. Specifically, what constitutes the flow equations? Do these theories feature both types of flow equations? These are among the significant questions that will be explored in our forthcoming studies.

\section*{Acknowledgments}
The authors are grateful to Dmitri Sorokin and Roberto Tateo for valuable discussions. S. H. acknowledges financial support from the Max Planck Partner Group, the Fundamental Research Funds for the Central Universities, and the Natural Science Foundation of China Grants No. 12075101 and No. 12235016. H. O. is supported by the National Natural Science Foundation of China, Grant No. 12205115, and by the Science and Technology Development Plan Project of Jilin Province of China, Grant No. 20240101326JC.

\begin{appendix}
\section{Details of Lagrangians}
In Section \ref{02}, we addressed the differential equation \ref{lagrangeNGZ} utilizing the ansatz ${\cal L}(\lambda^n)$  as delineated in Eq. \ref{lagrangeGeneral}, considering terms up to $ \lambda^7$, which led to the derivation of Lagrangian \ref{Ltz}. By emulating the methodology outlined in this section, the Lagrangian \ref{lagrangeGeneral} can be extrapolated to include terms up to $\lambda^{12}$, presented below:
\begin{eqnarray}
\label{Ap1}
{\cal L}({\lambda}^{12})&=&- t + \tfrac{1}{2} \lambda (t^2 + z^2) -  \tfrac{1}{2} t \lambda^2 (t^2 + z^2)+ \lambda^3 \bigl(\tfrac{1}{2} a_{11} (t^2 + z^2)^2 + \tfrac{1}{4} (t^4 -  z^4)\bigr)\\
&& + \lambda^4 \bigl(-2 a_{11} t (t^2 + z^2)^2 + \tfrac{1}{8} t (t^2 + z^2) (5 t^2 + 9 z^2)\bigr) \nonumber\\
&& + \lambda^5 \bigl(\tfrac{1}{3} a_{22} (t^2 + z^2)^3 + \tfrac{5}{3} a_{11} (t^6 - 3 t^2 z^4 - 2 z^6) + \tfrac{1}{6} (-4 t^6 + 15 t^2 z^4 + 11 z^6)\bigr)\nonumber\\
&& + \lambda^6 \bigl(-2 a_{11}^2 t (t^2 + z^2)^3 - 2 a_{22} t (t^2 + z^2)^3 + a_{11} t (t^2 + z^2)^2 (13 t^2 + 23 z^2)\nonumber\\
&& -  \tfrac{1}{16} t (t^2 + z^2) (101 t^4 + 286 t^2 z^2 + 193 z^4)\bigr)\nonumber\\
&&+ \lambda^7 \bigl(\tfrac{25}{4} t^8 -  \tfrac{189}{4} t^4 z^4 - 70 t^2 z^6 - 29 z^8 + \tfrac{7}{2} a_{11}^2 (t^2 - 3 z^2) (t^2 + z^2)^3\nonumber\\
&& + \tfrac{7}{4} a_{22} (t^2 - 3 z^2) (t^2 + z^2)^3 + \tfrac{1}{4} a_{37} (t^2 + z^2)^4 -  \tfrac{7}{2} a_{11} (t^2 + z^2)^2 (4 t^4 - 8 t^2 z^2 - 17 z^4)\bigr)\nonumber\\
&& +\lambda^8 \Bigl(-2 t a_{37}{} (t^2 + z^2)^4 + t{ a^2}_{11} (t^2 + z^2)^3 (71 t^2 + 127 z^2) + \tfrac{1}{3} t a_{22}{} (t^2 + z^2)^3 (79 t^2 + 135 z^2)\nonumber\\
&&  + \tfrac{1}{384} (51719 t^9 + 244356 t^7 z^2 + 429234 t^5 z^4 + 332468 t^3 z^6 + 95871 t z^8) + a_{11}{} \bigl(-4 t a_{22}{} (t^2 + z^2)^4\nonumber\\
&&  -  \tfrac{1}{12} t (t^2 + z^2)^2 (3469 t^4 + 9514 t^2 z^2 + 6381 z^4)\bigr)\Bigr) \nonumber\\
&&+ \lambda^9 \Bigl(- \tfrac{646}{5} t^{10} + 1533 t^6 z^4 + 3333 t^4 z^6 + \tfrac{5427}{2} t^2 z^8 + \tfrac{7843}{10} z^{10} + \tfrac{12}{5} a^3_{11} (t^2 - 4 z^2) (t^2 + z^2)^4\nonumber\\
&&  + \tfrac{9}{5} a_{37}{} (t^2 - 4 z^2) (t^2 + z^2)^4 + \tfrac{1}{5} a_{56}{} (t^2 + z^2)^5 -  \tfrac{6}{5} a_{22}{} (t^2 + z^2)^3 (22 t^4 - 66 t^2 z^2 - 123 z^4)\nonumber\\
&&  -  \tfrac{24}{5} a^2_{11} (t^2 + z^2)^3 (19 t^4 - 57 t^2 z^2 - 111 z^4) + a_{11}{} \bigl(\tfrac{36}{5} a_{22}{} (t^2 - 4 z^2) (t^2 + z^2)^4 \nonumber\\
&& + \tfrac{6}{5} (t^2 + z^2)^2 (241 t^6 - 482 t^4 z^2 - 2142 t^2 z^4 - 1454 z^6)\bigr)\Bigr) \nonumber\\
&&+ \lambda^{10} \Bigl(-2 t a^2_{22} (t^2 + z^2)^5 - 2 t a_{56}{} (t^2 + z^2)^5 + 15 t a_{37}{} (t^2 + z^2)^4 (3 t^2 + 5 z^2) + 4 t a^3_{11} (t^2 + z^2)^4 (37 t^2 + 67 z^2)\nonumber\\
&&-  \tfrac{1}{4} t a_{22}{} (t^2 + z^2)^3 (3535 t^4 + 9458 t^2 z^2 + 6259 z^4) -  \tfrac{3}{4} t a^2_{11} (t^2 + z^2)^3 (4841 t^4 + 13182 t^2 z^2 + 8901 z^4) \nonumber\\
&& -  \tfrac{1}{256} t (1177375 t^{10} + 6671235 t^8 z^2 + 15057590 t^6 z^4 + 16928966 t^4 z^6 + 9483355 t^2 z^8 + 2118119 z^{10})\nonumber\\
&&  + a_{11}{} \bigl(-4 t a_{37}{} (t^2 + z^2)^5 + 2 t a_{22}{} (t^2 + z^2)^4 (119 t^2 + 209 z^2)\nonumber\\
&&  + \tfrac{1}{8} t (t^2 + z^2)^2 (83861 t^6 + 308313 t^4 z^2 + 375363 t^2 z^4 + 151391 z^6)\bigr)\Bigr) \nonumber\\
&&+ \lambda^{11} \Bigl(\tfrac{52669}{12} t^{12} - 74877 t^8 z^4 -  \tfrac{638572}{3} t^6 z^6 - 255200 t^4 z^8 - 145002 t^2 z^{10} -  \tfrac{386057}{12} z^{12}  \nonumber\\
&&+ \tfrac{11}{3} a^2_{22} (t^2 - 5 z^2) (t^2 + z^2)^5 + \tfrac{11}{6} a_{56}{} (t^2 - 5 z^2) (t^2 + z^2)^5 + \tfrac{1}{6} a_{79}{} (t^2 + z^2)^6 \nonumber\\
&& -  \tfrac{11}{2} a_{37}{} (t^2 + z^2)^4 (8 t^4 - 32 t^2 z^2 - 55 z^4) - 132 a^3_{11} (t^2 + z^2)^4 (2 t^4 - 8 t^2 z^2 - 15 z^4) \nonumber\\
&& + \tfrac{11}{6} a_{22}{} (t^2 + z^2)^3 (466 t^6 - 1398 t^4 z^2 - 5157 t^2 z^4 - 3377 z^6) \nonumber\\
&& + a^2_{11} \bigl(\tfrac{22}{3} a_{22}{} (t^2 - 5 z^2) (t^2 + z^2)^5 + \tfrac{11}{3} (t^2 + z^2)^3 (1082 t^6 - 3246 t^4 z^2 - 12267 t^2 z^4 - 8191 z^6)\bigr) \nonumber\\
&& + a_{11}{} \bigl(\tfrac{22}{3} a_{37}{} (t^2 - 5 z^2) (t^2 + z^2)^5 -  \tfrac{220}{3} a_{22}{} (t^2 + z^2)^4 (4 t^4 - 16 t^2 z^2 - 29 z^4)  \nonumber\\
&&-  \tfrac{11}{6} (t^2 + z^2)^2 (5584 t^8 - 11168 t^6 z^2 - 78825 t^4 z^4 - 103226 t^2 z^6 - 41198 z^8)\bigr)\Bigr) \nonumber
\end{eqnarray}
\begin{eqnarray}
&&+ \lambda^{12} \Bigl(-2 t a_{79}{} (t^2 + z^2)^6 + 2 t a^4_{11} (t^2 + z^2)^5 (53 t^2 + 97 z^2) + t a_{56}{} (t^2 + z^2)^5 (69 t^2 + 113 z^2)  \nonumber\\
&&+ t a^2_{22} (t^2 + z^2)^5 (175 t^2 + 307 z^2) -  \tfrac{1}{4} t a_{37}{} (t^2 + z^2)^4 (8719 t^4 + 22982 t^2 z^2 + 15055 z^4) \nonumber\\
&& - 2 t a^3_{11} (t^2 + z^2)^4 (9865 t^4 + 26858 t^2 z^2 + 18313 z^4) + \tfrac{1}{1024} (234045265 t^{13} + 1551998034 t^{11} z^2 \nonumber\\
&& + 4275812203 t^9 z^4 + 6266390860 t^7 z^6+ 5153675607 t^5 z^8 + 2255709378 t^3 z^{10} + 410565197 t z^{12})\nonumber\\
&& + a_{22}{} \bigl(-4 t a_{37}{} (t^2 + z^2)^6  + \tfrac{1}{8} t (t^2 + z^2)^3 (354445 t^6 + 1288395 t^4 z^2 + 1552263 t^2 z^4 + 620425 z^6)\bigr)\nonumber\\
&&  + a^2_{11} \bigl(4 t a_{22}{} (t^2 + z^2)^5 (167 t^2 + 299 z^2)+ \tfrac{1}{8} t (t^2 + z^2)^3 (1849763 t^6 + 6772005 t^4 z^2 + 8234697 t^2 z^4 + 3327239 z^6)\bigr)\nonumber\\
&& + a_{11}{} \bigl(-4 t a_{56}{} (t^2 + z^2)^6 + 6 t a_{37}{} (t^2 + z^2)^5 (61 t^2 + 105 z^2) -  \tfrac{3}{2} t a_{22}{} (t^2 + z^2)^4 (11449 t^4 + 30730 t^2 z^2 + 20601 z^4)\nonumber\\
&&-  \tfrac{1}{64} t (t^2 + z^2)^2 (34854107 t^8 + 161619708 t^6 z^2 + 279804834 t^4 z^4 + 214485180 t^2 z^6 + 61452987 z^8)\bigr)\Bigr)\nonumber\,.
\end{eqnarray}
The derivation of the energy-momentum tensor from Lagrangian \ref{Ltzxalpa} incorporates terms to the order of $\lambda^4$. Due to its extensive nature, we present the $\lambda^4$ order of this energy-momentum tensor in Eq. \ref{Tmunuroot} as follows:
\begin{eqnarray}\label{O4}
T_{\mu \nu }(\lambda,\gamma )&\simeq& \lambda^4 \Biggl(\frac{\bigl(\cosh(\alpha -  \gamma)\bigr)^5 \sech (\gamma) d_{29}{} z^4 \bigl(5 \sech (\alpha) T^{Max}_{\mu \nu } \tanh(\alpha -  \gamma) + 4 z \mathit{g}_{\mu \nu }\bigr)}{-1 + 2 \cosh(2 \gamma) - 2 \cosh(4 \gamma)} \\
&&-  \frac{3 \bigl(\cosh(\alpha -  \gamma)\bigr)^2 \sech (\alpha) \bigl(\sech (\gamma)\bigr)^7 (d_{5}{})^2 d_{12}{} z^4 }{8 \bigl(1 - 2 \cosh(2 \gamma)\bigr)^2 \cosh(4 \gamma) \bigl(1 - 2 \cosh(2 \gamma) + 2 \cosh(4 \gamma)\bigr)}\times \biggl(-  \Bigl( 9 \sinh(\alpha - 7 \gamma)-9 \sinh(\alpha - 13 \gamma) \nonumber\\
&& + 39 \sinh(\alpha - 5 \gamma) + 55 \sinh(3 \alpha - 5 \gamma) + 46 \sinh(\alpha - 3 \gamma)+ 57 \sinh(\alpha + \gamma) - 25 \sinh\bigl(3 (\alpha + \gamma)\bigr)  \nonumber\\
&&+ 25 \sinh(3 \alpha + \gamma) + 34 \sinh(\alpha + 3 \gamma) - 55 \sinh\bigl(3 (\alpha + 3 \gamma)\bigr) + 14 \sinh(\alpha + 5 \gamma) + 2 \sinh(\alpha + 11 \gamma)\Bigr)  T^{Max}_{\mu \nu }\nonumber\\
&&+ 2 \Bigl(3 \cosh(2 \alpha - 13 \gamma) - 3 \cosh(2 \alpha - 7 \gamma) - 24 \cosh(2 \alpha - 5 \gamma) - 11 \cosh(4 \alpha - 5 \gamma) - 30 \cosh(2 \alpha - 3 \gamma) \nonumber\\
&&- 19 \cosh(\gamma) - 48 \cosh(3 \gamma) - 11 \cosh(5 \gamma) - 3 \cosh(7 \gamma) + 14 \cosh(11 \gamma) + 3 \cosh(13 \gamma) - 24 \cosh(2 \alpha + \gamma)\nonumber\\
&& - 5 \cosh(4 \alpha + \gamma) + 22 \cosh(\alpha) \cosh\bigl(3 (\alpha + 3 \gamma)\bigr) - 13 \cosh(2 \alpha + 3 \gamma) \nonumber\\
&&+ 5 \cosh(4 \alpha + 3 \gamma) + 2 \cosh(2 \alpha + 5 \gamma) + 14 \cosh(2 \alpha + 11 \gamma)\Bigr) z \mathit{g}_{\mu \nu }\biggr)\nonumber\\
&&+ \frac{9 \bigl(\cosh(\alpha -  \gamma)\bigr)^3 \sech (\alpha) \bigl(\sech (\gamma)\bigr)^3 (d_{12}{})^2 z^4 }{4 \bigl(1 - 2 \cosh(2 \gamma)\bigr)^2 \bigl(1 - 2 \cosh(2 \gamma) + 2 \cosh(4 \gamma)\bigr)}\times \biggl(\bigl(-3 \cosh(5 \gamma) + 5 \cosh(2 \alpha + 3 \gamma)\bigr) T^{Max}_{\mu \nu } \nonumber\\
&&+ 2 \Bigl(- \sinh(\alpha - 5 \gamma) + \sinh\bigl(3 (\alpha + \gamma)\bigr) + 2 \cosh(\gamma) \sinh(\alpha + 4 \gamma)\Bigr) z \mathit{g}_{\mu \nu }\biggr)\nonumber\\
&&+ \frac{4 \bigl(\cosh(\alpha -  \gamma)\bigr)^3 \sech (\alpha) \bigl(\sech (\gamma)\bigr)^3 d_{5}{} d_{14}{} z^4 }{\cosh(4 \gamma) \bigl(1 - 2 \cosh(2 \gamma) + 2 \cosh(4 \gamma)\bigr)}\times\biggl(\bigl(-3 \cosh(5 \gamma) + 5 \cosh(2 \alpha + 3 \gamma)\bigr) T^{Max}_{\mu \nu } \nonumber\\
&&+ 2 \Bigl(- \sinh(\alpha - 5 \gamma) + \sinh\bigl(3 (\alpha + \gamma)\bigr) + 2 \cosh(\gamma) \sinh(\alpha + 4 \gamma)\Bigr) z \mathit{g}_{\mu \nu }\biggr)\nonumber\\
&& -  \frac{\cosh(\alpha -  \gamma) \bigl(\sech (\gamma)\bigr)^{11} (d_{5}{})^4 z^4 }{8 \bigl(1 - 2 \cosh(2 \gamma)\bigr)^2 \cosh(4 \gamma) \bigl(1 - 2 \cosh(2 \gamma) + 2 \cosh(4 \gamma)\bigr)}\times \Bigl(\bigl(3 \cosh(2 \alpha - 17 \gamma) - 21 \cosh(2 \alpha - 11 \gamma)\nonumber\\
&& - 69 \cosh(2 \alpha - 9 \gamma) - 55 \cosh(4 \alpha - 9 \gamma) - 107 \cosh(2 \alpha - 7 \gamma) + 3 \cosh(2 \alpha - 5 \gamma) - 21 \cosh(2 \alpha - 3 \gamma)\nonumber\\
&& + 55 \cosh(4 \alpha - 3 \gamma) + 89 \cosh(2 \alpha -  \gamma) + 65 \cosh(4 \alpha -  \gamma) - 19 \cosh(\gamma) - 17 \cosh(3 \gamma) + 7 \cosh(5 \gamma) \nonumber\\
&&- 17 \cosh(7 \gamma) - 19 \cosh(9 \gamma) + 7 \cosh(15 \gamma) + 61 \cosh(2 \alpha + \gamma) + 75 \cosh(2 \alpha + 3 \gamma) + 65 \cosh(4 \alpha + 3 \gamma)\nonumber\\
&& + 133 \cosh(2 \alpha + 5 \gamma) + 55 \cosh(4 \alpha + 5 \gamma) + 89 \cosh(2 \alpha + 7 \gamma) - 55 \cosh(4 \alpha + 11 \gamma)\nonumber\\
&& - 35 \cosh(2 \alpha + 13 \gamma)\bigr) \sech (\alpha) T^{Max}_{\mu \nu } + 4 \bigl(\sinh(2 \alpha - 17 \gamma) - 7 \sinh(2 \alpha - 11 \gamma) - 23 \sinh(2 \alpha - 9 \gamma) \nonumber\\
&&- 11 \sinh(4 \alpha - 9 \gamma) - 43 \sinh(2 \alpha - 7 \gamma) + \sinh(2 \alpha - 5 \gamma) - 7 \sinh(2 \alpha - 3 \gamma) + 11 \sinh(4 \alpha - 3 \gamma) \nonumber\\
&&+ 37 \sinh(2 \alpha -  \gamma) + 13 \sinh(4 \alpha -  \gamma) + 33 \sinh(\gamma) + 15 \sinh(3 \gamma) + 57 \sinh(5 \gamma) + 63 \sinh(7 \gamma) \nonumber\\
&& - 9 \sinh(15 \gamma)+ 33 \sinh(9 \gamma) + 29 \sinh(2 \alpha + \gamma) + 25 \sinh(2 \alpha + 3 \gamma) + 13 \sinh(4 \alpha + 3 \gamma) + 53 \sinh(2 \alpha + 5 \gamma)\nonumber\\
&& + 11 \sinh(4 \alpha + 5 \gamma) + 37 \sinh(2 \alpha + 7 \gamma) - 11 \sinh(4 \alpha + 11 \gamma) - 19 \sinh(2 \alpha + 13 \gamma)\bigr) z \mathit{g}_{\mu \nu }\Bigr)\Biggr)\,.\nonumber
\end{eqnarray}

\end{appendix}

%%%%%%%%%%%%%%%%%%%%%%%%%%%%%%%%%%%
%%%%%%%%%%%%%%%%%%%%%%%%%%%%%%%%%%%%%%%%%%%%%%%%%%%%%%%%%%%%%%%%%%%%%%%%%%%%%%%%%%%%%%%%%%%%%%%%%%%%%%%%%%%%%%%%%%%%%%%%%%%%%%%%%%%%%%%%%%%%%%%%%%%%%%%%%%%%%%%%%%%%%%%%%%%%%%%%%%%%%%%%%%%%%%%%%%%%%%%%%%%%%%%%%%%%%%%%%%%%%%%%%%%%%%%%%%%%%%%%%%%%%%%%%%%%%%%%%%%%%%%%%%%%%%%%%%%%%%%%%%%%%%%%%%%%%%%%%%%%%%%%%%%%%

\if{}
\bibliographystyle{abe}
\bibliography{references}{}
\fi

\providecommand{\href}[2]{#2}\begingroup\raggedright\endgroup

\end{document}